\documentclass{pasj00}
\draft

\begin{document}
\SetRunningHead{Markowitz et al.}{{\it Suzaku} observation of NGC 3516}
\Received{2007/12/31}
\Accepted{2007/12/31}

\title{The {\it Suzaku} Observation of NGC 3516: Complex 
Absorption and the Broad and Narrow Fe K Lines}


\author{
Alex \textsc{Markowitz}\altaffilmark{1,2},
James N. \textsc{Reeves}\altaffilmark{1,2,3},
Giovanni \textsc{Miniutti}\altaffilmark{4},
Peter \textsc{Serlemitsos}\altaffilmark{1},
Hideyo \textsc{Kunieda}\altaffilmark{5,6},
Tahir \textsc{Yaqoob}\altaffilmark{2},
Andrew C. \textsc{Fabian}\altaffilmark{4},
Yasushi \textsc{Fukazawa}\altaffilmark{7},
Richard \textsc{Mushotzky}\altaffilmark{1},
Takashi \textsc{Okajima}\altaffilmark{1,2},
Luigi C. \textsc{Gallo}\altaffilmark{8},
Hisamitsu \textsc{Awaki}\altaffilmark{9},
Richard E. \textsc{Griffiths}\altaffilmark{10}}
\altaffiltext{1}{X-ray Astrophysics Laboratory, Code 662, NASA/Goddard Space Flight Center, Greenbelt, MD 20771, USA}
\altaffiltext{2}{Department of Physics and Astronomy, Johns Hopkins University, Baltimore, MD 21218, USA}
\altaffiltext{3}{Astrophysics Group, School of Physical and Geographical Sciences, Keele University, Keele, Staffordshire ST5 5BG, UK}
\altaffiltext{4}{Institute of Astronomy, Madingley Road, Cambridge CB3 0HA, UK}
\altaffiltext{5}{Department of Astrophysics, School of Science, Nagoya University, Chikusa-ku, Nagoya, 464-01, Japan}
\altaffiltext{6}{Institute of Space and Astronautical Science, JAXA, Sagamihara, Kanagawa 229-8510, Japan}
\altaffiltext{7}{Department of Physics, Hiroshima University, 1-3-1 Kagamiyama, Higashi-Hiroshima, 739-8526, Japan}
\altaffiltext{8}{SUPA, School of Physics and Astronomy, University of St.\ Andrews, North Haugh, St.\ Andrews, KY16 9SS, UK}
\altaffiltext{9}{Department of Physics, Faculty of Science, Ehime University, Bunkyo-cho, Matsuyama, Ehime, 790-8577, Japan} 
\altaffiltext{10}{Department of Physics, Carnegie Mellon University, 5000 Forbes Avenue, Pittsburgh, PA 15213, USA}


\maketitle

\begin{abstract}       
We present results from a 150 ksec {\it Suzaku} observation of the Seyfert 1.5 NGC 3516 
in October 2005. The source was in a relatively highly absorbed state. Our best-fit 
model is consistent with the presence of a low-ionization absorber which has a column 
density near 5 $\times$ 10$^{22}$ cm$^{-2}$ and covers most of the X-ray continuum source
(covering fraction 96--100$\%$).
A high-ionization absorbing component, which yields a narrow absorption feature consistent with
Fe K {\sc XXVI}, is confirmed.
A relativistically broadened Fe K$\alpha$ line is required 
in all fits, even after the complex absorption is taken into account; 
an additional partial-covering component is an inadequate substitute for the continuum curvature associated with the broad Fe line. 
A narrow Fe K$\alpha$ emission line 
has a velocity width consistent with the Broad Line Region. The 
low-ionization absorber may be responsible for producing the narrow Fe K$\alpha$ line, though
a contribution from additional material out of the line of sight is possible. 
We include in our model soft band emission lines from He- and H-like ions of N, O, Ne and Mg, 
consistent with photo-ionization, though a small 
contribution from collisionally-ionized emission is possible.
\end{abstract}

\section{Introduction}

The 6.4 keV Fe K$\alpha$ emission line has long been known to 
be an important diagnostic of the material accreting onto supermassive black holes.
The Compton reflection hump, frequently seen in Seyfert spectra above 
$\sim$7~keV and peaking near 20--30 keV (Pounds et al.\ 1990), indicates that Seyferts' 
Fe K lines may have an origin in optically thick material, such as the accretion disk.
Observations with {\it ASCA} indicated that many Fe K$\alpha$ lines were broad 
(FWHM velocities up to $\sim$0.3$c$) and asymmetrically skewed towards lower 
energies, implying an origin in the inner accretion disk; the line profile is
sculpted by gravitational redshifting and relativistic Doppler effects (e.g., 
Tanaka et al.\ 1995, Fabian et al.\ 2002). However, {\it XMM-Newton} and {\it Chandra} 
observations have been revealing a more complex picture. A narrow Fe K component 
(FWHM velocities $\sim$5000 km s$^{-1}$ or less) appears to be quite 
common; these lines' widths suggest emission from distant material, such as the 
outer accretion disk, the optical/UV Broad Line Region (BLR) or the molecular torus.
Spectral observations in which the broad and narrow components are deconvolved 
are thus a prerequisite for using the Fe K line as a tracer of the geometry of 
the emitting gas.

At the same time, there is strong evidence from X-ray and UV grating observations 
for the presence of ionized material in the inner regions of a large fraction of 
AGN (e.g., Blustin et al.\ 2005; McKernan et al.\ 2007). High-resolution spectroscopy shows the gas is 
usually outflowing from the nucleus; typical velocities are $\sim$ a few hundred 
km s$^{-1}$. Absorption due to a broad range of ionic species is commonly seen;
and for many sources, there is evidence for several different photo-ionized
absorbing components, as opposed to a single absorber, along the line of sight. 
In the Fe K bandpass, some Seyferts show evidence for absorption by H- or He-like 
Fe, indicating a zone of highly-ionized absorbing material (e.g., NGC 3783, Reeves 
et al.\ 2004).

Cold absorbing gas, with line of sight columns in excess of the Galactic value, 
are routinely observed in Seyfert 2 AGN in accordance with unification schemes (Urry, 
Padovani 1995), and have also been reported in some Seyfert 1 AGN. Importantly, 
variations in column density on timescales from hours to years have been observed 
in both Seyfert 1 AGN (e.g., in I Zw 1, Gallo et al.\ 2007;  
see also Lamer et al.\ 2003, Puccetti et al.\ 2004) and Seyfert 2 
AGN (Risaliti et al.\ 2002; Risaliti et al.\ 2005), suggesting that the 
absorbing circumnuclear material is not homogeneous in either Seyfert type, has a 
high tranverse velocity and occurs over a range of length scales.

NGC 3516 is a well-studied, nearby (z = 0.008836; Keel 1996) Seyfert 1.5 AGN that 
can display strong 2--10 keV flux variability on timescales of hours to years 
(e.g., Markowitz, Edelson 2004). Previous X-ray spectroscopic observations of 
NGC 3516 e.g., Nandra et al.\ (1997) using {\it ASCA}, have indicated the presence 
of a broad Fe K line, and this source is known to also contain complex and ionized 
absorption. Numerous UV absorption lines, including N {\sc V}, C {\sc IV} and 
Si {\sc IV}, were observed with the {\it International Ultraviolet Explorer} 
(Ulrich, Boisson 1983); absorption line strengths vary on timescales as short 
as weeks as the absorber responds to variations in the ionizing flux (e.g., Voit et al.\
1987). {\it Hubble Space Telescope} observations have revealed 
that this component of absorbing gas (henceforth called the ``UV absorber'')
may consist of several distinct kinematic components (Crenshaw et al.\ 1998).

X-ray spectra of NGC 3516 can exhibit evidence for large columns 
($\gtrsim$10$^{22}$ cm$^{-2}$) of absorbing gas (e.g., Kolman et al.\ 1993), and
the X-ray absorbers can display variability on timescales of years (e.g., 
Mathur et al.\ 1997). Using {\it Chandra} gratings data from observations 
in April 2001 and November 2001, Turner et al.\ (2005) observed K-shell absorption lines
due to H- like Mg, Si and S, and He-like Si, evidence for a highly-ionized 
absorber, likely with column density $\gtrsim$10$^{22}$ cm$^{-2}$, outflowing 
at $\sim$1100 km s$^{-1}$. Simultaneous with these {\it Chandra} observations 
in 2001 were two {\it XMM-Newton} observations. Turner et al.\ (2005) modeled 
the continuum curvature of the two {\it XMM-Newton} EPIC spectra by including 
a partial covering, mildly-ionized absorber; the column density was 
$\sim$2.5 $\times$ 10$^{23}$ cm$^{-2}$, with a covering fraction of $\sim$50$\%$.
However, the formal requirement for the broad Fe line was reduced, leading to 
uncertainty as to whether the broad Fe line really existed in NGC 3516.
Spectral fitting using an instrument with a wide bandpass is thus necessary 
to remove such model degeneracies.

In this paper, we report on an observation of NGC 3516 made with the {\it Suzaku} 
observatory in October 2005. The combination of the X-ray Imaging Spectrometer 
(XIS) CCD and the Hard X-ray Detector (HXD) instruments have yielded a broadband 
spectrum covering 0.3 to 76 keV, allowing us to deconvolve the various broadband 
emitting and absorbing components. Furthermore, the exceptional response of the 
XIS CCD and high signal-to-noise ratio of this observation have allowed us to 
study narrow emission lines in great detail. $\S$2 gives a brief overview of the 
{\it Suzaku} observatory, and describes the observation and data reduction. $\S$3 
describes fits to the time-averaged spectrum. Variability analysis is briefly
discussed in $\S$4. Flux-resolved spectral fits are discussed
in $\S$5. In $\S$6, we describe a search for narrow red- or blue-shifted lines
in the Fe K bandpass.
The results are discussed in $\S$7, and a brief summary is given in $\S$8.

\section{Observations and Data Reduction}

The nucleus of NGC 3516 was observed by {\it Suzaku} from 2005 October 12 at 
13:45 UT until October 15 at 09:07 UT. {\it Suzaku} was launched 2005 July 10 
into a low-Earth orbit. It has four X-ray telescopes (XRTs; Serlemitsos et 
al.\ 2007), each with a spatial resolution of 2$\arcmin$ (HPD). The XRTs focus 
X-rays onto four X-ray Imaging Spectrometer (XIS; Koyama et al.\ 2007) CCDs, 
which are sensitive to 0.2--12 keV X-rays on a 18$\arcmin$ by 18$\arcmin$ 
field of view, contain 1024 by 1024 pixel rows each, and feature an energy 
resolution of $\sim$150 eV at 6 keV. Three CCDs (XIS0, 2 and 3) are 
front-illuminated (FI), the fourth (XIS1) is back-illuminated (BI) and features
an enhanced soft X-ray response. The XRT/XIS combination yields effective areas 
per detector of roughly 330 cm$^{2}$ (FI) or 370 cm$^{2}$ (BI) at 1.5 keV,
and 160 cm$^{2}$ (FI) or 110 cm$^{2}$ (BI) at 8 keV. Each XIS is equipped with 
two $^{55}$Fe calibration sources which produce fluorescent Mn K$\alpha$ and 
K$\beta$ lines and are located at the CCD corners. {\it Suzaku} also features 
a non-imaging, collimated Hard X-ray Detector (HXD; Takahashi et al.\ 2007); 
its two detectors, PIN and GSO, combine to yield sensitivity from $\sim$10 
to $\sim$700 keV. Further details of the {\it Suzaku} observatory are given in
Mitsuda et al.\ (2007).

\subsection{XIS Reduction}

The XIS data used in this paper were version 1.2 of the screened data (Fujimoto 
et al.\ 2007) provided by the Suzaku team. The screening is based on the following 
criteria: grade 0, 2, 3, 4, and 6 events were used, the {\sc cleansis} script was 
used to remove hot or flickering pixels, data collected within 256 s of passage 
through the South Atlantic Anomaly (SAA) were discarded, and data were selected 
to be 5$\arcdeg$ in elevation above the Earth rim (20$\arcdeg$ above the day-Earth 
rim). The XIS-FI CCDs were in 3x3 and 5x5 editmodes, for a net exposure time after 
screening of 135.0 (XIS0), 134.8 (XIS2) and 135.2 (XIS3) ks. XIS1 was also in 3x3 
and 5x5 editmodes, for a net exposure of 135.4 ks. The XIS was in normal clocking mode.

The source was observed at the nominal center position of the XIS. For each XIS, 
we extracted a 3$\arcmin$ radius centered on the source. The background was 
extracted using four circles of radius 1.5$\arcmin$, each located $\sim$6$\arcmin$ 
from the source, but chosen to avoid the z = 2.1 QSO RX J110741.4+723235, located 
4.5$\arcmin$ SE of NGC 3516. Spectra were binned to a minimum of 25 counts 
bin$^{-1}$ to allow use of the $\chi^2$ statistic.

Response matrices and ancillary response files (ARFs) were generated for each 
XIS independently using {\sc xissimrmfgen} and {\sc xissimarfgen} version 
2006-10-26 (Ishisaki et al.\ 2007). The ARF generator takes into account the 
level of hydrocarbon contamination on the optical blocking filter. We estimate 
a carbon column density of  0.8, 1.2, 1.7, and 2.8 $\times$ 10$^{18}$ cm$^{-2}$ 
for XIS0, 1, 2 and 3, respectively. Finally, we co-added the three XIS-FI spectra 
using {\sc mathpha}, and co-added the response files and ARFs using {\sc addrmf} 
and {\sc addarf}, respectively.


To examine the accuracy of the XIS RMFs and determine residual line width due 
e.g., to imperfect CTI correction, we generated spectra for the $^{55}$Fe 
calibration source lines on each XIS using the above response matrices and ARFs. 
We fit the calibration source spectra with three Gaussians. Two Gaussians were 
for the Mn K$\alpha$ doublet (expected energies 5.899 keV and 5.888 keV), with 
energy centroids fixed to be 11 eV apart, and the higher energy line flux set 
to twice that of the lower energy one. The third Gaussian was used to model the 
K$\beta$ line, expected at 6.490 keV.  We found the average of all the calibration 
line widths $\sigma$ to be $<$ 4 eV. The Mn K$\alpha$ line energy centroids for 
the co-added FI spectrum were consistent with the expected energies to within 
1 eV. For XIS1, the Mn K$\alpha$ line energy centroids were 3$\pm$2 eV lower 
than expected. Such discrepancies are well within the accuracy ($\sim$0.2\% at 
Mn-K$\alpha$) of the energy calibration of XIS. Fitting the calibration source 
lines without the response file, we determined the FWHM energy resolution during
the observation to be 145 eV (average of the 4 XISes).

\subsection{HXD Reduction}

We used data from the HXD-PIN only; NGC 3516 was extremely faint in the HXD-GSO 
band, and some aspects of the GSO background are still being studied, so we defer
analysis of the GSO data to a later time. The PIN source spectra were extracted 
from cleaned version 1.2 (pre-1.2-r1) HXD event files provided by the HXD 
instrument team. PIN background count rates are variable and strongly depend on 
the time since SAA passage (Kokubun et al.\ 2007), so we selected data according 
to the following criteria: at least 500 s since SAA passage, cutoff rigidity (COR) $\geq$ 8 GV,
and day- and night-Earth elevation angles each $\geq$5$\arcdeg$. Instrumental 
(non-X-ray) background spectra for the PIN were provided by the HXD Team 
(``Background A'' model) generated from a time-dependent model. The model utilized 
the count rate of upper discriminators as the measure of cosmic ray flux that 
passed through the silicon PIN diode and yielded background spectra based on a 
database of non X-ray background observations with the PIN (Fukazawa et al.\ 2007). 
The current accuracy of the PIN non-X-ray background (NXB) model 
for a 1 day observation is about 5$\%$ 
(peak-to-peak residuals).
Both the source and background spectra were generated with identical good time 
intervals, and the source exposure was corrected for instrument dead time 
(a $\sim$5$\%$ effect). This yielded a good time exposure of 105.8 ks.

Data $<$ 12 keV were discarded due to noise contamination near the lower 
threshold of the PIN diode. Data above 76 keV were also discarded: the gain 
above an internal Bi K$\alpha$ calibration line at 76 keV is not well-defined, 
though the photon statistics above this energy were poor anyway for this 
observation. Further details of the HXD in-orbit performance are given in 
Kokubun et al.\ (2007). To model the contribution to the total background from 
the Cosmic X-ray Background (CXB), a spectrum of the form 
9.0 $\times$ 10$^{-9}$($E$/3keV)$^{-0.29}$ exp($-E$/40keV) erg cm$^{-2}$ 
s$^{-1}$ sr$^{-1}$ keV$^{-1}$ (Gruber et al.\ 1999) was used. We note that some 
recent works (e.g., Frontera et al.\ 2007) suggest a 10$\%$ normalization increase 
compared to Gruber et al.\ (1999).  In addition, 
spatial fluctuations of order $\sim$5--10$\%$
over $\sim$ a square degree are
known (e.g., Barcons et al.\ 1998).
However, the effect on the net spectrum was 
negligible; for instance, the change in Compton reflection component strength 
was 1$\%$, typically. To simulate a CXB spectrum using {\sc xspec}, we assumed 
a model of the form {\sc powerlaw*highecut}\footnote{The {\sc highecut} model is of the form $I(E) = exp(  (E_{\rm c}-E)/E_{\rm f})$ for $E$ $>$ $E_{\rm c}$,
and $I(E) = 1$ for $E$ $<$ $E_{\rm c}$, where $E$ is the photon energy, $E_{\rm c}$ is the cutoff energy, and $E_{\rm f}$ is the e-folding energy.}, 
with photon index $\Gamma$ = 1.29, cutoff energy of 0.1 keV, and an e-folding energy of 40 keV. 
The power-law normalization of 8.8 $\times$ 10$^{-4}$ ph cm$^{-2}$ s$^{-1}$ keV$^{-1}$ (at 1 keV)
was used, appropriate for a source observed in XIS-nominal mode. The 12--76 
keV CXB flux was 1.4 $\times$ 10$^{-11}$ erg cm$^{-2}$ s$^{-1}$ (using the 
Gruber et al.\ 1999 normalization). The total (X-ray plus particle) background 
12--76 keV flux was 4.4 $\times$ 10$^{-10}$ erg cm$^{-2}$ s$^{-1}$.

The spectrum was binned to a minimum of 400 count bin$^{-1}$. We used the 
response file ae$\_$hxd$\_$pinxinom$\_$20060814.rsp. The mean 12--76 keV 
net source flux and count rate were 1.1 $\times$ 10$^{-10}$ erg cm$^{-2}$ s$^{-1}$ 
and 0.16 ct s$^{-1}$, respectively. Figure 1 shows the net source, background, and total 
(source + background) spectra. The source spectrum is always at least 15$\%$ 
of the total up to $\sim$40 keV.

\section{Model Fits to the Time-Averaged Spectrum}

We used 0.4--11.5 keV data in the XIS-FI spectrum and 0.3--10.0 keV data in 
the XIS-BI spectrum. We ignored 1.72--1.87 keV in the co-added FI spectrum and
1.80--1.87 keV in the BI spectrum due to uncertainties in calibration associated 
with the instrumental Si K edge. In all fits, we included a constant to account 
for relative instrument normalizations. We left the relative XIS-BI/XIS-FI 
normalization free, but best-fit values were always extremely close to 1.00.
The PIN/XIS-FI normalization was kept fixed at 1.13, a value derived using 
{\it Suzaku} observations of the Crab (Ishida et al.\ 2006); the uncertainty
on the PIN/XIS-FI normalization is also discussed in $\S$3.1. All errors 
on one interesting parameter correspond to $\Delta\chi^2 = 2.71$ (with the 
XIS-BI/XIS-FI normalization left free). 
All fits were done using {\sc xspec} v.11.3.2.
The abundances of Lodders (2003) were used.  A neutral 
Galactic column of 2.94 $\times$ 10$^{20}$ cm$^{-2}$ was included (Dickey, Lockman 1990).

\subsection{Preliminary Broadband Fits}

The X-ray continuum emission of Seyfert 1 and 1.5 AGN is usually dominated by a 
power-law component thought to originate from Comptonization of soft 
seed photons by a hot corona near the central black hole (e.g., Shapiro et al.\
1976;  Sunyaev, Titarchuk 1980; Haardt et al.\ 1994). A 
simple power-law (henceforth denoted the ``primary power-law'') over 
0.3--76 keV yielded a very poor fit, with $\chi^2$/$dof$ (degrees of freedom) = 
49708/1449.  As shown in Figure 2(a), residuals strongly indicated the need 
to include absorption to account for continuum curvature $\lesssim$3--4 keV. The 
narrow 6.4 keV Fe K$\alpha$ line was also obvious.

We therefore added an absorbing column to the power-law, assuming systemic redshift 
and initially assuming a covering fraction of unity (this component is henceforth denoted the 
``primary absorber''). We used {\sc xstar} tables that assumed an underlying continuum
with a photon index of $\Gamma$ = 2.0 to model absorption in this paper.
We also added Gaussians to model Fe K$\alpha$ and K$\beta$ emission lines. The 
Fe K$\beta$ energy centroid was kept fixed at 7.06 keV; the normalization was
kept fixed at 0.13 times that of the K$\alpha$ line. With these changes to 
the model, $\chi^2$/$dof$ fell to 8312/1444. The best-fit ionization 
parameter and column density were log($\xi$) = 2.0$\pm$0.1 erg cm s$^{-1}$ and 
7.1$\pm$0.2 $\times$ 10$^{22}$ cm$^{-2}$, respectively ($\xi$ $\equiv$ 4$\pi$$F_{\rm ion}$/$n$;
$F_{\rm ion}$ is the 1--1000 Ryd ionizing continuum flux; $n$ is the density 
of the reflecting material). However, as shown in Figure 2(b), this model did 
not accurately describe the broadband emission.

To attempt to model the soft emission, we added a second power-law (the 
``soft power-law''), with photon index $\Gamma$ tied to the primary power-law;
$\chi^2$/$dof$ fell to 3039/1443.  
As shown in Figure 2(c), residuals suggested the presence of soft
X-ray emission lines, e.g., near 0.56 keV, which is likely due to O {\sc VII}.

We added 13 Gaussians to our model, widths $\sigma$ were fixed at 0.5 eV.
Energy centroids for the lower signal-to-noise ratio lines were kept fixed 
at lab-frame energies. Table 1 shows the results for the lines in our best-fit 
baseline model (see below). Data/model residuals for these emission lines are 
shown in Figure 3. Removing lines one at a time from the final fit suggested that it was 
significant at $\geq$ 99.0$\%$ confidence in an $F$-test 
to include each line in the fit. However we caution that
the resulting $F$-test significance levels
were likely upper limits in a few cases, since
there could be some blending of Gaussian profiles when two lines' energy 
centroids are relatively close together.
In addition, changing the order 
in which lines were added might also affect
the derived significance of any one line.\footnote{For further discussion regarding justification of
including multiple lines in complex spectral fits,
we refer the reader to Pounds \& Vaughan (2006), who demonstrate
an application of a Bayesian analysis technique.}
We identify these lines as originating in H-like 
C, N, O, Ne and Mg, and He-like N, O, Ne and Mg. We also report radiative 
recombination continua (RRC) of O {\sc VII}, O {\sc VIII} and Ne {\sc IX} (and 
possibly C {\sc VI}, blended with the 0.500 keV N {\sc VII} emission line).
The lines are likely due to photo-ionization.
Most of these lines have been reported previously; we refer the reader to
Turner et al.\ (2003) for results using the {\it XMM-Newton}-RGS. The strongest 
line detected in both the RGS spectrum and the {\it Suzaku} spectrum is due to
O {\sc VII}; line intensities of N {\sc VI}, N {\sc VII}/C {\sc VI} RRC blend, O {\sc VII}, and O {\sc VIII} 
as measured by {\it Suzaku} were roughly consistent with the RGS measurements.
We also included a line near 0.83 keV for Fe L {\sc XVII}, to 
model any contribution from collisionally-ionized gas. Turner et al.\ (2003) 
included a Mg {\sc XI} recombination edge component. However, we eschewed it in favor of 
a Mg {\sc XII} line at 1.47 keV; 
the higher effective area of the XIS compared to that of the RGS at this energy allowed us to
determine that a line gave a better fit than a recombination edge. Including these emission lines in the model 
resulted in $\chi^2$/$dof$ falling to 2347/1424. Figure 2(d) shows the 
data/model residuals after the soft X-ray lines are modeled.


Residuals in the PIN band, peaking near 20--30 keV, signaled the need to include
a Compton reflection component. We modeled this by adding a {\sc pexrav} component
(Magdziarz, Zdziarski 1995), assuming 
solar abundances and an input photon index tied to that of the primary power-law.
We fixed the cut-off energy at 400 keV. This choice was somewhat arbitrary, but
the cutoff energy is not well constrained; in the baseline model below, 
we found a lower limit of 120 keV. We initially fixed the inclination
at 30$\arcdeg$, as per NGC 3516's classification as a Seyfert 1.5, and
we initially assumed that the reflector is subject to the same absorption as the primary 
power-law. In the best fit-model, $\chi^2$/$dof$ fell to 1895/1423, and the 
value of the reflection fraction $R$ (defined as $\Omega$/2$\pi$, where $\Omega$ 
is the solid angle subtended by the reflector) was 2.8$^{+0.4}_{-0.2}$. The 
photon index $\Gamma$ was 2.142$^{+0.020}_{-0.007}$. As shown in Figure 2(e), 
the residuals in the PIN band were thus corrected. 

The uncertainty on the relative PIN/XIS cross-normalization is about 3\% (Ishida et al.\ 2006); the 
subsequent effect on $R$ is smaller than that associated with the uncertainty 
of the PIN NXB. We modified the intensity of the PIN NXB by $\pm$2\%, which 
is the 1$\sigma$ level of the current reproducibility of the PIN NXB. $R$ 
changed by $\pm$0.2, which is smaller than the statistical error on $R$.
Errors quoted on $R$ for the remainder of this paper are the statistical 
errors only; readers should bear in mind the additional systematic uncertainty 
associated with the NXB.

There remained residuals in the 5--6 keV band. We first discuss modeling these 
residuals using relativistic diskline components, and later ($\S$3.2) we will discuss
if a partial covering component could provide as good a fit. We added two relativistic diskline
components for Fe K$\alpha$ and Fe K$\beta$ emission, using
a Kerr (maximally rotating) black hole line profile (Laor 1991).
Formally, the K$\beta$ diskline is not required (omission of this component does 
not change $\chi^2$/$dof$ significantly), but we include it for completeness.
The normalization of the K$\beta$ diskline was fixed at 0.13 times that of the
K$\alpha$ diskline. The Fe K$\alpha$ line energy was constrained to lie within 
rest-frame energies of 6.40 and 6.96 keV; the K$\beta$ line energy was fixed at 
7.06 keV. All other parameters were kept tied between the K$\alpha$ and K$\beta$ 
components.  The emissivity index $\beta$ (used when quantifying the radial 
emissivity per unit area as a power-law, r$^{-\beta}$) was fixed at 3.0. Initially, 
the disk inclination $i$ was fixed at 30$\arcdeg$. The outer radius $R_{\rm out}$ was 
kept fixed at 400 $R_{\rm g}$ (1 $R_{\rm g}$ $\equiv$ G$M_{\rm BH}$/$c^2$). The 
inner radius $R_{\rm in}$ was left free. With 
the best-fit model, $\chi^2$/$dof$ fell to 1643/1420 ($\Delta\chi^2$ = --252), and the data/model 
residuals near 5--6 keV were improved considerably; see Figure 2(f). 

Refitting with the diskline inclination $i$ as a free parameter
yielded a significant improvement in the fit:
$\chi^2$/$dof$ fell to 1521/1419 ($\Delta\chi^2$ = --122) for $i$ $<$ 23$\arcdeg$,
significant at $>$99.99$\%$ confidence in an $F$-test.
In this and all subsequent best-fit models, we fixed the inclination 
of the Compton reflection component to match that of the diskline as 
opposed to leaving it fixed at 30$\arcdeg$, though in practice this usually had negligible impact
on the fit. 
We refer to this model as the ``1-absorber + Compton reflection + diskline'' model.
Additional model parameters are listed in Table 2.


Visual inspection of the data/model residuals in the Fe K bandpass suggested an  
additional dip near 6.9 keV, at the rest-frame energy for Fe {\sc XXVI}; 
see Figure 4(a). H- and He-like Fe K absorption features might be expected, 
given the detection of a high-ionization absorber by Turner et al.\ (2005). 
Adding a Gaussian at 6.96 keV, $\chi^2$ fell by 7.9 (for one less $dof$), 
significant at 99.4$\%$ in an $F$-test.
The absolute values of the intensity and equivalent width $EW$
relative to the primary power-law were 3$\pm$2 $\times$ 10$^{-6}$ ph cm$^{-2}$ s$^{-1}$ and 9$\pm$5 eV.
Adding a narrow ($\sigma$ = 0.5 eV) 
Gaussian with energy centroid fixed at 6.70 keV yielded an improvement in fit of 
$\Delta\chi^2$ = --6.3, significant at 98.4$\%$ in an $F$-test. 
The absolute values of the intensity and equivalent width $EW$ were 
4$\pm$2 $\times$ 10$^{-6}$ ph cm$^{-2}$ s$^{-1}$ and 9$\pm$4 eV, respectively.

The {\it Chandra}-HETGS spectrum yielded narrow absorption features in the 1--3 keV band due to
highly-ionized Mg, Si, and S, but we do not significantly detect any narrow
absorption features at those energies (and we would likely not expect to, 
given the XIS resolution).  A small dip in the spectrum near 2.3 keV 
is close to the expected energy for S {\sc XII}. However, {\sc xstar} models 
demonstrate that S {\sc XV} absorption, though not significantly detected here, 
is always stronger than S {\sc XII}. This feature is more likely due to 
calibration uncertainty associated with an instrumental Au M edge. 
 

To model Fe K absorption features, we added a second absorbing component 
(henceforth denoted the ``high-ionization absorber''), again using an {\sc xstar} table. 
Based on the results of 
Turner et al.\ (2005), we assumed an outflow velocity of 1100 km s$^{-1}$. 
$\chi^2$/$dof$ fell to 1485/1417 ($\Delta\chi^2$ = --36). In this model, the high-ionization absorber 
had a column 4.0$^{+4.6}_{-3.1}$ $\times$ 10$^{22}$ cm$^{-2}$, similar to the value 
used by Turner et al.\ (2005), and log($\xi$) = 3.7$^{+0.3}_{-0.7}$ erg cm s$^{-1}$.
We henceforth refer to this model as our  ``2-absorber'' or ``baseline'' model. 
Data/model residuals are shown in Figures 2(g) and 4(b); the 6.9 keV
residuals are reduced and there is some slight improvement to the $>$7 keV continuum.
The primary power-law, 
with $\Gamma$ = 1.91$^{+0.04}_{-0.05}$, was absorbed by a column 
5.5$\pm$0.2 $\times$ 10$^{22}$ cm$^{-2}$ and log($\xi$) = 0.3$\pm$0.1 erg cm s$^{-1}$. 
Forcing the ionization parameter to a much lower value (e.g., log($\xi$) $\sim$ --0.5) 
resulted in a significantly worse fit, with large residuals at 1.5--3.0 keV; this is 
likely because the lower ionization does not account for absorption edges due to 
higher ionization species of Si and S. We caution, however, that this result 
could be influenced by residual calibration effects in the XIS near 1.8 keV and 2.3 keV 
(instrumental Si K and Au M edges, respectively). 
Refitting the model with the diskline emissivity as a free parameter
yielded no additional improvement to the fit. We found $\beta$ = 3.2$\pm$0.3; 
we leave $\beta$ frozen at 3.0 in subsequent fits.
The values of $EW$ for the broad and narrow lines were 287$^{+49}_{-24}$ and
123$\pm$7 eV, respectively. Figure 4(c) shows the data/model 
residuals when the broad lines were removed from the model. When the model was then re-fit,
there were correlated residuals in the Fe K bandpass, as illustrated in Figure 4(d).
In addition, the resulting high 
value of $\chi^2$/$dof$ (1651/1421; $\Delta\chi^2$ = +166) compared to the baseline model indicated 
that removing the disklines yielded a significantly worse fit ($>$99.999$\%$ 
in an $F$-test).\footnote{We note that when removing the PIN data, and fitting using only the XIS,
the uncertainty on the broad line flux increases by 
20$\%$, and the uncertainty on $\Gamma$ increases by 80$\%$.}

In the best-fit baseline model, the strength of the Compton reflection hump was
$R$ = 1.7$^{+0.4}_{-0.5}$; a contour plot of $R$ as a function of $\Gamma$ is 
shown in Figure 5. Also shown in Figure 5 is a contour plot of $R$ as a function 
of the PIN/XIS-FI normalization, which had been kept frozen at 1.13; in this plot, 
$\Gamma$ was a free parameter. Figure 6 shows an unfolded model spectrum. Other 
model parameters are listed in Table 2 (see Table 1 for the soft X-ray emission lines).

Finally, we re-fit the model, assuming that the Compton reflection component was
not affected by the primary absorber. This yielded a goodness of fit nearly 
identical to the previous fit, with $\chi^2$/$dof$ = 1479/1417.
All fit parameters were virtually identical to the previous fit; for instance,
$R$ was 1.9$\pm$0.4; we will continue to assume that the Compton reflection component 
is absorbed.

We returned to focus on the soft X-ray emission lines, and attempted 
to model all emission lines using an {\sc xstar} grid for photo-ionized 
line emission; the grid assumed an underlying photon index of 2.0. We 
also included an extra narrow Gaussian at 0.84$\pm$0.02 keV, again 
likely associated with Fe L {\sc XVII} emission from collisionally-ionized 
gas. Assuming a single photo-ionized emission zone with abundances for 
N, O, Ne, Mg and Fe left as free parameters yielded $\chi^2$/$dof$ = 
1559/1424 for values of the ionization parameter log($\xi$) = 
1.3$^{+0.1}_{-0.2}$ erg cm s$^{-1}$ and column density 1.4$^{+0.3}_{-0.5}$ 
$\times$ 10$^{23}$ cm$^{-2}$. Best-fit abundance values relative to 
solar were $Z_{\rm N}$ = 1.2$^{+1.2}_{-0.3}$, $Z_{\rm O}$ = 0.5$^{+0.3}_{-0.1}$,
$Z_{\rm Ne}$ = 1.0$^{+0.4}_{-0.2}$, $Z_{\rm Mg}$ = 2.8$\pm$0.6, and 
$Z_{\rm Fe}$ = 0.3$\pm$0.3, i.e., there is a similar indication of 
a high value of N/O as inferred by Turner et al.\ (2003) from 
the {\it XMM-Newton} RGS spectrum. Adding a second zone of emission from
ionized gas with a significantly different ionization parameter did 
not improve the fit. While the fit using the one-zone model is not 
poor, for the remainder of this paper, we will continue to use the
multiple Gaussians to model the photo-ionized emission lines, as 
that model yielded a lower value of $\chi^2$/$dof$.

\subsection{Additional Partial Covering Components}

We next explored the possibility of an additional, partial covering component
whose presence could potentially be manifested as curvature in the 1--5 keV continuum.
We differentiate such components from the soft power-law, which itself could 
indicate ``leaked'' emission as part of a partial covering scenario (see $\S$7.1).
Starting with the baseline model, we added a partial covering component 
consisting of a power-law (with photon index tied to that of the primary 
power-law) absorbed by an {\sc xstar} component. We first kept the column 
density $N_{\rm H, PC}$ and ionization parameter log($\xi_{\rm PC}$) tied 
to those of the primary absorber. In the best-fit model (henceforth denoted 
Model PC1), the added power-law had a normalization 0.32$\pm$0.03 times that 
of the primary power-law. As shown in Table 3, $N_{\rm H}$ and log($\xi$) of 
the primary absorber and the diskline parameters were consistent with values 
of the baseline model. However, $\chi^2$/$dof$ was 1486/1416, virtually 
identical to the baseline model, and so there is no formal requirement to 
include the new partial covering component when the column density and 
ionization states are tied to those for the primary absorber. 


Next, we untied $N_{\rm H, PC}$ and log($\xi_{\rm PC}$) and refit (Model PC2). 
The best-fit parameter values are listed in Table 3. Compared to Model PC1,
$\chi^2$ dropped by only 5.2 for 2 additional degrees of freedom, significant
at only $\sim$90$\%$ in an $F$-test. It is thus not highly significant to include
the new partial covering component in this case either.


Next, we addressed whether it was possible for a partial covering component
to mimic the curvature in the $\sim$4--6 keV continuum modeled above as a 
relativistically-broadened Fe line. Starting with the baseline model, we 
removed the disklines, and added a partial covering component consisting of 
a power-law (with photon index tied to that of the primary power-law)
absorbed by gas with a relatively low value of 
the ionization parameter (Model PC3). The column density needed to be 
$\gtrsim$10$^{22.5}$ cm$^{-2}$ in order to have a rollover near 2--3 keV,
causing an apparent ``peak'' near $\sim$4--6 keV
(we forced it to be greater than 4 $\times$ 10$^{22}$ cm$^{-2}$).
In the best-fit model, the partial coverer had a column density of 
7.8$^{+68}_{-3.8}$ $\times$ 10$^{22}$ cm$^{-2}$
(error pegged at lower limit), and 
log($\xi_{\rm PC}$) = --0.4$^{+1.2}_{-2.6}$ (error pegged at lower limit). 
The new partially-covered power-law had a normalization $<$0.22 
times that of the primary power-law. However, $\chi^2$/$dof$ was 1677/1418 ($\Delta\chi^2$ = +192), 
a much worse fit compared to the baseline model, and there were still large 
data/model residuals in the Fe K bandpass; see Figure 2(h). We conclude that 
a partial coverer cannot mimic the observed curvature of the diskline.

\subsection{Relativistic Reflection Fits}

The broad Fe K diskline component is a signature of reflection off a possibly-ionized disk.
{\it Suzaku}'s broad bandpass makes it an ideal instrument for attempting to 
model the entire reflection spectrum (broad Fe line plus hard X-ray reflection 
continuum plus soft X-ray emission) of Seyferts in a self-consistent manner. 
Specifically, we used the ionized reflection models of Ross, Fabian (2005) modified by 
relativistic smearing.

We first removed the disklines and Compton reflection components from 
the baseline model, and modeled a blurred reflector by convolving
a Ross, Fabian (2005) reflection 
model with the kernel associated with a maximally rotating Kerr black hole line profile model 
(Laor 1991) using {\sc kdblur}. 
The photon index of the illuminating continuum was tied to that of the primary power-law.
Free parameters here were the emissivity index $\beta$, inner disk 
radius $R_{\rm in}$ and disk inclination $i$. 
We also included the narrow Fe K line at 6.4 keV; 
this model assumes that all of the observed Compton reflection is associated with the broadened Fe K line
and none with the narrow Fe K line. The best-fit model had $\chi^2$/$dof$ = 1486/1416, similar to the best overall fit;
the results suggested a fairly neutral or lowly-ionized reflector, with log($\xi$) $<$ 1.8 erg cm s$^{-1}$. 
The best-fit value of $R$ was 3.0$\pm$0.8; the photon index $\Gamma$ was 2.06$\pm$0.08.  
Parameters associated with blurring were similar to what was 
obtained in the baseline model: an inclination of 26$\pm$7$\arcdeg$ and
an inner disk radius $<$ 5.5 $R_{\rm g}$ (the emissivity index was
fixed at 3.0). The absorber parameters were consistent to those obtained for the baseline model.

Next, we tested a model with both a blurred reflector and an unblurred reflector,
to test the notion that both the broad and narrow lines each be associated with contributions
to the observed Compton reflection continuum. 
The blurred reflector was again modeled by convolving a Ross, Fabian (2005) reflection model
with with a relativistic diskline profile. 
The unblurred reflector consisted of a narrow 6.4 keV line and a Compton reflection hump
modeled using {\sc pexrav}. The fit was very
similar to the previous one and to the baseline model: $\chi^2$/$dof$ = 1485/1415.
The strengths of the blurred and unblurred Compton reflection components
were 1.8$\pm$0.7 and 1.6$\pm$0.5, respectively. All other parameters were
consistent with those of the previous fit (e.g., the $EW$ of the narrow Fe K line
was identical to that measured in the baseline model).

\subsection{Narrow Fe K Line Properties}

In our baseline model, the best-fit energy centroid for the narrow
Fe K$\alpha$ line was 6.398$\pm$0.004 keV, consistent with neutral Fe.
The observed line width $\sigma_{\rm obs}$ was 26$^{+11}_{-13}$ eV.
The intrinsic line width $\sigma_{\rm intr}$ was found by subtracting 
in quadrature the $^{55}$Fe calibration line width $\sigma$ of $<$4 eV
from the measured line width.  We inferred a 99$\%$ confidence limit for
two interesting parameters of $\sigma_{\rm intr}$ $<$ 45 
eV, which corresponds to a FWHM velocity width $<$4900  km s$^{-1}$.
Figure 7 shows a contour plot of line intensity versus FWHM velocity width.
This width is consistent with the results obtained by {\it Chandra}-HETGS for the
two observations in 2001, 1290$^{+1620}_{-1290}$ and 3630$^{+2350}_{-1540}$ km s$^{-1}$ 
(Yaqoob, Padmanabhan 2004).

Figure 8 shows a contour plot of the broad line intensity versus the
narrow line intensity, illustrating that the two lines are detected 
independently at $>$4$\sigma$ confidence. Such a result is a product of 
the combination of the narrow response of the XIS (yielding extremely 
high signal/noise in the narrow line) and {\it Suzaku}'s broad bandpass.
Similar results have been reported e.g., for the {\it Suzaku} observation of
NGC 2992 (Yaqoob et al.\ 2007).

Finally, we discuss limits to a Fe K$\alpha$ Compton shoulder. We added 
a Gaussian emission line at 6.24 keV (rest-frame), with width tied to 
that of the K$\alpha$ core. We found an upper limit to the intensity of 
7 $\times$ 10$^{-6}$ ph cm$^{-2}$ s$^{-1}$, or 13$\%$ of the K$\alpha$ 
core intensity. This limit corresponds to an $EW$ of 21 eV.

\section{Timing Analysis}

To compare the variability properties of the primary and soft power-laws,
we extracted light curves, summed over all four XISes, orbitally-binned, 
and background-subtracted, for the 0.3--1.0 and 2--10 keV bands. 
The 2--10 keV light curve is plotted in Figure 9, along with the 12---76 keV 
background-subtracted PIN light curve, binned every three orbits. 
The 0.3--1.0 keV light curve was consistent with being constant within the errors
and is not plotted.
We note that the PIN errors shown in Figure 9 are statistical only, and do not take into
account systematic uncertainty associated with subtraction of the NXB component. 
For instance, Mizuno et al.\ (2006) noted that the systematic error 
is roughly 6$\%$ for the 15--40 keV band over a 5760 s bin. 
We calculated the fractional variability amplitude $F_{\rm var}$ 
(which quantifies the variability in excess of measurement noise) and 
its uncertainty following Vaughan et al.\ (2003). For the 2--10 keV band, 
$F_{\rm var}$ was $9.2 \pm 0.3 \%$, with a maximum/minimum flux ratio
of roughly 1.4. For the 12--76 keV band, no significant variability in excess of that
due to measurement errors was detected, with an upper limit of 4.4$\%$. 
The upper limit on $F_{\rm var}$ for the 0.3--1.0 keV band was 2.5$\%$.

\section{Flux-resolved Spectral Fits}

We performed flux-resolved spectral fits to search for any physical 
connection between the soft and primary power-laws. Despite the limited 
flux range exhibited during this observation, we attempted to determine, 
e.g., if the observed X-ray flux variability could be due to rapid variations 
in column density of the primary absorber, if both power-laws vary together, 
or if one power-law is constant. We split the time-averaged spectrum into 
periods when the 2--10 keV flux was higher and lower than the average 2--10 
keV flux of  2.31 $\times$ 10$^{-11}$ erg cm$^{-2}$ s$^{-1}$, as illustrated 
in Figure 9. Net exposure times for the high and low-flux spectra for each 
XIS (and the PIN) were approximately 62.1 (46.3) and 71.3 (59.5) ks, respectively.
The average 2--10 keV fluxes were 2.47 and 2.16 $\times$ 10$^{-11}$ erg cm$^{-2}$ s$^{-1}$, 
respectively. 
We applied the best-fitting baseline model from the time-average spectrum to 
both spectra. All narrow Gaussian energy centroids and widths, $\beta$, 
$i$ and $R_{\rm in}$ for the disklines, and log($\xi$) for the high-ionization 
absorber were kept frozen at their time-averaged values. However, all freed 
parameters (including $\Gamma$ and $R$) were consistent at the 90$\%$ confidence level. For instance, 
we find no strong evidence that the column density of the high-ionization 
absorber varies on short timescales. A broadband observation of NGC 3516 spanning a 
larger flux range is thus needed to potentially distinguish determine if 
the two power-laws vary in concert.

\section{The Search for Red and Blue-shifted Narrow Lines}

We searched for additional narrow absorption or emission features in the 
Fe K bandpass, as seen so far in several Seyferts, including NGC 3516.
For instance, Turner et al.\ (2002) found emission lines near 5.57, 6.22, 
6.41, 6.53 and 6.9 keV in the November 2001 {\it Chandra}-HETGS and 
{\it XMM-Newton} EPIC spectra of NGC 3516. In the April 2001 observation,
Iwasawa et al.\ (2004) claimed the presence of a transient feature with varying 
energy and with flux varying on timescales of 25 ks. One interpretation was that 
such features were red- and blue-shifted Fe K lines associated
with transient ``hot spot'' emission on the inner accretion disk.

We searched for such features by adding a Gaussian component to the time-averaged
spectrum, with width frozen at 0.5 eV, sliding it over 4--9 keV in energy.
In addition to the aforementioned absorption line at 6.96 keV,
there were only two ``candidate'' feature with $\Delta\chi^2$ $<$ --5.0,
absorption lines near 6.0 and 6.7 keV. However, we performed Monte Carlo 
simulations to assess the statistical significance of these features 
(see $\S$3.3  of Porquet et al.\ 2004 and $\S$4.3.3 of Markowitz et al.\
2006 for a description of these simulations, also Gallo et al.\ 2005), and
we found that the lines were consistent with photon noise. Furthermore, 
each candidate features was evident only in one XIS camera, and thus 
was likely not real.

As an aside, we note that a 7.47 keV Ni K$\alpha$ emission line was not detected
in the time-average spectrum; adding a Gaussian at this energy,
we found an upper limit of 3 eV.

We also extracted time-resolved spectral slices by dividing the time-averaged 
spectrum into five slices 48 ks in duration, with each slice having an exposure 
time near 27 ks per XIS and 21 ks for the PIN. We did not investigate longer time
slices since they might miss short-lived hot-spot emission lines; shorter time 
slices would have yielded poorer photon statistics. Applying a sliding narrow 
Gaussian over 4--9 keV in each spectral slice revealed only 3 ``candidate'' 
emission or absorption lines (with $\Delta\chi^2$ $<$ --6). Again, however, the 
features were seen in only one or two XIS cameras, and Monte Carlo simulations 
showed that not a single candidate feature was inconsistent with photon noise
at greater than 80$\%$ confidence. Analysis of the high-and low-flux spectra similarly
yielded no significant narrow emission or absorption lines (even at 6.70 and 6.96 keV). 
Typical upper limits to the equivalent width of an emission line at 5.57 keV 
(one of the energies of the transient lines in Turner et al.\ 2002), were
$\lesssim$10--15 eV in the time-averaged spectrum or any of the sub-spectra.

\section{Discussion}

During the late 1990's, NGC 3516 typically displayed a 2--10 keV flux of 
$\sim$4--6 $\times$ 10$^{-11}$ erg cm$^{-2}$ s$^{-1}$ (e.g., Markowitz, Edelson 2004).
During the 2001 {\it XMM-Newton}/{\it Chandra} observations, however, the observed 
2--10 keV flux was much lower: 1.6--2.3 $\times$ 10$^{-11}$ erg cm$^{-2}$ s$^{-1}$ 
(Turner et al.\ 2005). Table 4 lists the inferred absorption-corrected 2--10 keV nuclear 
fluxes from the {\it XMM-Newton} observations, as well as during the 2005 {\it Suzaku} 
observation. In addition, Figure 10 shows the unfolded observed spectra for the {\it Suzaku} XIS and the
2001 {\it XMM-Newton} EPIC-pn data (see Turner et al.\ 2005 for details regarding the
{\it XMM-Newton} data). The {\it Suzaku} observation apparently caught the source in a similar 
low level of nuclear flux as the 2001 observations.  The observed 0.5--2.0 keV flux 
during the {\it Suzaku} observation, however, was $\sim$2--3 times lower than during 
the {\it XMM-Newton} observations, indicating that the source was still heavily 
obscured, and confirming that the complex absorption in this source cannot be ignored 
when fitting the broadband spectrum and modeling diskline components.

\subsection{Power-law Components}


The primary power-law observed in the hard X-rays is likely emission from a hot corona 
very close to the supermassive black hole, as seen in all Seyferts. 
The nature of the
soft power-law component, however, is not as clear. It could represent nuclear emission
scattered off optically-thin material, e.g., in the optical/UV Narrow Line Region (NLR). 
In the baseline model, the normalization of the soft power-law relative to that 
of the primary power-law was 4.2$\pm$0.4$\%$. Assuming a covering fraction of unity,
this ratio is equal to the optical depth of the scattering material, 
indicating a column density of roughly 5 $\times$ 10$^{22}$ cm$^{-2}$, 
consistent with this notion, though the column density is somewhat too high to 
likely be associated with the NLR.  
It is interesting to note that this column density is similar to that obtained
for the high-ionization absorber, suggesting
the possibility that this absorbing component could be associated with a zone of scattering.
Using {\it Chandra}-ACIS, George 
et al.\ (2002) found the extended circumnuclear gas to have a 0.4--2.0 keV flux of 
roughly 10$^{-14}$ erg cm$^{-2}$ s$^{-1}$. However, that emission was studied over an 
annular extraction region 3$\arcsec$ to 10$\arcsec$ (0.6--1.8 kpc), and so that flux value is likely a 
lower limit to the 0.4--2.0 keV flux that Suzaku would observe (given the XRTs' 2$\arcmin$ HPD). In our baseline model, 
we found an unabsorbed 0.4--2.0 keV flux of 1 $\times$ 10$^{-12}$ erg cm$^{-2}$ s$^{-1}$, consistent with 
the notion that the soft power-law is scattered emission.

In this case, the decrease in observed 0.5--2.0 keV flux from 2001 to 2005 could potentially
be explained by the scattered emission responding to a recent decrease in nuclear continuum flux.
However, this scenario would require the bulk of the scattered emission to lie within at most
a few light years of the black hole, and the nuclear flux would have had to decrease between 2001 and 2005
(when the source was not observed by any major X-ray mission in the 2--10 keV band) then return
to 2001 levels by the {\it Suzaku} observation.


Alternatively, the soft power-law could be unobscured, ``leaked'' nuclear 
emission as part of a partial covering scenario. In this case, the primary absorber 
would obscure 96$\%$ of the sky as seen from the nuclear continuum source. The lack 
of significant variability in the 0.3--1.0 keV band could argue for the soft power-law to 
originate in scattered emission, since we might expect to observed variability of the 
same amplitude as the 2--10 keV band only if the soft power-law were leaked nuclear 
emission. However, this is far from certain, as the 0.3--1.0 keV band had a low count rate
and the presence of the soft emission lines 
in the XIS spectrum could contribute to dilution of intrinsic variability in the 
soft power-law. A broadband observation spanning a larger observed flux range is 
needed to clarify this issue. The soft power-law could of course represent a blend of 
scattered emission plus leaked nuclear emission. We therefore conclude that the primary 
absorber has a covering fraction between 96--100$\%$.

\subsection{Complex Absorption}

We detect two zones of absorption: in addition to the
primary absorber, which has a covering fraction of 
96--100$\%$, there is the high-ionization absorber, which is
assumed here to have a covering fraction 
of unity. The high-ionization absorber is potentially the same as that reported by Turner et al.\ (2005);
we find a column density $N_{\rm H}$ of 4.0$^{+4.6}_{-3.1}$ $\times$ 10$^{22}$ cm$^{-2}$,    
consistent with the column density of 2 $\times$ 10$^{22}$ cm$^{-2}$ used by Turner et al.\ (2005), although we use a slightly higher 
ionization parameter (see Turner et al.\ 2007).
Previous studies of NGC 3516, such as Netzer et al.\ (2002), have discussed in detail 
the UV absorber, responsible for H Ly$\alpha$, C {\sc IV} and N {\sc V} absorption 
features in {\it Hubble Space Telescope} spectra (Kraemer et al.\ 2002). In the X-ray band, 
discrete features associated with Mg {\sc VII--IX} and Si {\sc VII--IX} are expected from
this component, but with the CCD resolution and with calibration-related artifacts
near 1.7--1.8 keV in the XIS, such features are not detected by {\it Suzaku}.

{\it Suzaku} has found the primary absorber of the hard X-ray continuum to be
lowly-ionized (log($\xi$) = 0.3$\pm$0.1 erg cm s$^{-1}$), with a column density 
$N_{\rm H}$ of 5.5$\pm$0.2 $\times$ 10$^{22}$ cm$^{-2}$. It is possible that it is 
the same absorber that Turner et al.\ (2005) designated as the ``heavy'' 
partial-covering absorber, though we use a somewhat lower ionization parameter (see Turner et al.\ 2007).
A new observation of NGC 3516 with {\it XMM-Newton} in 2006 October showed that
the source had returned to similar $>$6 keV brightness and similar
obscuration levels ($N_{\rm H}$ $\sim$ 2 $\times$ 10$^{23}$ cm$^{-2}$; covering fraction $\sim$ 45$\%$)
as during the 2001 observations (Turner et al.\ 2007). 
Long-term changes in the covering fraction of the heavy absorber could explain the bulk of
the spectral variability changes between the 2001, 2005 and 2006 observations. 
In this case, the column density has decreased by a 
factor of 4.5, while the covering fraction has increased 
from $\sim$40--60$\%$ to 96--100$\%$, from 2001 to 2005, 
and subsequently returned to approximately the same levels as in 2001 within the next 12 months.
However, because we have not actually observed the entire eclipse associated with a
specific, discrete blob of absorbing gas traversing the line of sight, it is not 
clear whether the covering fractions derived are associated with single, large blobs
partially blocking the line of sight to the X-ray continuum source, or if the absorber 
consists of numerous, discrete blobs or has a filamentary or patchy structure. 
On the other hand, given the gaps  
between the 2001 and 2006 {\it XMM-Newton} and 2005 {\it Suzaku} observations, it is certainly plausible 
that these observations could have caught independent, discrete blobs or filaments
with differing column densities and differing physical sizes and/or radial distances 
lying on the line of sight.

To estimate the distance $r$ between the central black hole and the absorbing 
gas, we can use a definition of the
ionization parameter $\xi$ = $L_{1-1000 {\rm Ryd}}$/($n$$r^2$), where $n$ is the number 
density. $L_{1-1000 {\rm Ryd}}$ is the 1--1000 Ryd illuminating continuum luminosity, 
and the value of the ionization parameter is taken to be the 
value of 2 erg cm s$^{-1}$ measured above. We estimate the maximum possible 
distance to the material by assuming that the thickness $\Delta$$r$ 
must be less than the distance $r$. The column density 
$N_{\rm H}$ = $n$$\Delta$$r$, yielding the upper limit 
$r$ $<$ $L_{1-1000 {\rm Ryd}}$/($N_{\rm H}$$\xi$). We estimate the 1-1000 Ryd
flux from the baseline model to be 9.9 $\times$ 10$^{-11}$ erg cm$^{-2}$ s$^{-1}$,  which 
corresponds to  $L_{1-1000 {\rm Ryd}} = 1.7 \times 10^{43}$ erg s$^{-1}$ (assuming 
$H_{\rm o}$ = 70 km s$^{-1}$ Mpc$^{-1}$ and $\Lambda_{\rm o}$ = 0.73).    
$r$ is thus $<$2 $\times$ 10$^{20}$ cm  (180 light-years), a very loose upper 
limit encompassing both distances associated with the BLR ($\sim$10 light-days; 
Peterson et al.\ 2004) and a possible cold molecular torus at a 1 pc radius. 
Variability in the absorber properties between 2001--2005 and
2005--2006 thus imply a radial distance of at most a few light years in the case
of clouds traversing the line of sight to the nucleus.
In addition, in the case of a partial covering scenario, it is plausible that
the absorber's size could be of the same order as that of the X-ray continuum 
source. The absorbing material could thus be e.g., associated with the
base of an outflow or from dense clouds associated with 
magnetohydrodynamic disk turbulence (e.g., Emmering et al.\ 1992).

NGC 3516's transition from an unobscured source to a moderately-obscured 
source in a 4 year span presents a challenge to standard Seyfert 1/2 unification 
schemes. If the obscuration in NGC 3516 is associated with an equatorial molecular torus
usually invoked in Seyfert 1/2 unification schemes, then it is possible that
during the {\it Suzaku} observation, the inner edge of the torus
could have intersected the line of sight, but given
NGC 3516's classification as a Seyfert 1, this could only occur if the torus opening angle were
extremely small and the torus were not azimuthally symmetric. 
Alternatively, variations in column density and/or covering fraction 
could be due to fine structure in large-distance (tens of pc), non-equatorial 
filaments that traverse the line of sight (e.g., Malkan et al.\ 1998).

\subsection{Fe K Emission Components and Compton Reflection}

We have deconvolved
the broadband emitting components, and determined that 1) the existence of the broad Fe line 
is robust in that it was required in all models for an adequate fit, and 2) 
a partial covering component could not mimic the curvature associated with 
a relativistic broad line. 
We note, for instance, that if we remove the diskline components from the baseline model and refit,
not only is the fit worse ($\chi^2$ increases by over 170), but the
value of $R$ becomes $\sim$ 3.2.  This value is incompatible with the observed $EW$ of the
narrow line unless the Fe abundance is extremely sub-solar ($\lesssim$0.3; see below).
The best-fit disk inclination was typically 
$\lesssim$25$\arcdeg$. The inner radius was typically $\lesssim$5$R_{\rm g}$.
The line energy was seen to be consistent with neutral to mildly-ionized Fe
(up to Fe $\sim$ {\sc XX}; Kallman et al.\ 2004).
The equivalent width with respect the primary continuum was 287$^{+49}_{-24}$ eV,
consistent with the value of 431$^{+193}_{-172}$ eV obtained by Turner et al.\ (2005) for the April 2001
{\it XMM-Newton} observation,
where the spectrum could be fit with a diskline component in addition to the 
complex absorbing components.

The line energy of the narrow Fe K$\alpha$ line was also consistent with emission from neutral Fe.
The intensity of the narrow line during the {\it Suzaku} observation
is roughly 40$\%$ higher than that measured during the
2001 {\it XMM-Newton} and {\it Chandra} observations (Turner et al.\ 2002), possibly indicating
that a substantial fraction of the Fe K line photons originate in a region $\lesssim$5 lt.-yr.\ in size.
We measured a FWHM velocity line width for the narrow Fe K$\alpha$ line of $<$ 4900
km s$^{-1}$ (99$\%$ confidence level for two interesting parameters). 
This velocity does not rule out an origin in the 
BLR; Peterson et al.\ (2004) reported FWHM velocities 
for the H$\alpha$ and H$\beta$ lines of 4770$\pm$893 and 3353$\pm$310 
km s$^{-1}$, respectively.  However, we also 
cannot exclude a contribution from an origin in the putative molecular torus; there could 
potentially be a very narrow line component with FWHM velocity $\sim$ a few hundred km s$^{-1}$,
but the XIS would be unable to separate it from the relatively broader line component.

It is possible that the same material that absorbs the hard X-rays along the line of sight is 
responsible for producing the narrow Fe line. 
The material producing the Fe line cannot have a column substantially less than 
10$^{\sim 22}$ cm$^{-2}$ or else there would be insufficient optical depth
to produce a prominent Fe K line. The primary absorber, with its column density of 
5.5 $\times$ 10$^{22}$ cm$^{-2}$ and low ionization state, is thus a plausible candidate for 
the narrow line origin.
As an estimate of the Fe K$\alpha$ equivalent 
width expected in this case, we can assume an origin in optically-thin gas 
which completely surrounds a single X-ray continuum source and is uniform 
in column density, and use the following equation:
\begin{equation}
EW_{\rm calc} = f_{\rm c} \omega f_{\rm K\alpha} A \frac{\int^{\infty}_{E_{\rm K edge}}P(E) \sigma_{\rm ph}(E) N_{\rm H} dE}{P(E_{\rm line})}
\end{equation}

Emission is assumed to be isotropic. Here, $f_{\rm c}$ is the covering 
fraction, initially assumed to be 1.0. $\omega$ is the fluorescent yield, 
0.34 (Kallman et al.\ 2004). $f_{\rm K\alpha}$ is the fraction of photons 
that go into the K$\alpha$ line as opposed to the K$\beta$ line; this is 
0.89 for Fe {\sc I}. $A$ is the number abundance relative to hydrogen. 
We assumed solar abundances, using Lodders (2003). $P(E)$ is the spectrum 
of the illuminating continuum at energy $E$; $E_{\rm line}$ is the 
K$\alpha$ emission line energy. $\sigma_{\rm ph}(E)$ is the photo-ionization 
cross section assuming absorption by K-shell electrons only
(Veigele 1973\footnote{http://www.pa.uky.edu/$\sim$verner/photo.html}).

For $N_{\rm H}$ = 5.5 $\times$ 10$^{22}$ cm$^{-2}$, $EW_{\rm calc}$ = 29 eV, 
substantially lower than the observed $EW$ of 123$\pm$7 eV. We conclude that it is possible for 
the primary absorber to contribute to the observed line $EW$, but there is also likely a 
contribution from some other (non-continuum absorbing) material lying out of the line of sight, likely
with column densities 10$^{\sim 23}$ cm$^{-2}$ (e.g., Matt et al.\ 2002).
For instance, if the putative cold molecular torus does not intersect the line of sight, it could 
contribute to the observed $EW$.  The 13$\%$ upper limit to ratio of the
Compton shoulder/ narrow Fe K$\alpha$ core intensity was a 90$\%$ confidence limit only,
and does not exclude at high confidence the possibility of Compton-thick material out of the line of sight.


An additional possibility is that
the material emitting the bulk of the line photons could be responding to
a continuum flux that was higher in the past. 
For instance, if the putative molecular torus is located $\sim$ a pc or so from the black hole,
the torus will yield a line $EW$ corresponding to the 
continuum flux averaged over the past few years.
This situation is plausible for NGC 3516, as the 2--10 keV flux of NGC 3516 
during $\sim$1998--2001 (Markowitz, Edelson 2004) 
was a factor of $\sim$1.5--2 times brighter than during 2005.

We now turn our attention to properties of the Compton reflection continuum.
{\it Suzaku} has observed other Seyferts to display reduced levels of variability 
in the PIN band compared to the 2--10 keV band, e.g., in MCG--6-30-15 (Miniutti et al.\ 
2007). This behavior is thought to be caused by the presence of the relatively non-varying
Compton reflection hump, which dilutes the observed $>$10 keV variability of the 
power-law component. Gravitational light-bending in the region of strong gravity has been invoked
to explain the relative constancy of the reflection spectrum (Compton hump and Fe K diskline component) despite
large variations in the coronal power-law flux in MCG--6-30-15, for instance (Miniutti et al.\ 2007).
In the case of NGC 3516, the observed fractional variability amplitudes
for the 2--10 and 12--76 keV bands were $F_{\rm var,2-10}$  = 
9.2$\pm$0.3$\%$ and $F_{\rm var,12-76}$ $<$ 4.4$\%$, respectively. These measurements allow us to
rule out the possibility that the Compton hump varies in concert with the power-law, since
the variability amplitudes would be consistent in that case. 
The primary power-law and Compton hump contribute 44$\%$ and 56$\%$, respectively, 
of the total 12--76 keV flux. In the case of a constant Compton hump and variable power-law, 
$F_{\rm var,12-76}$ would then be equal to $F_{\rm var,2-10}$ / 2.25, or roughly 4.1$\%$.
This is roughly consistent with the observed upper limit on $F_{\rm var,12-76}$,
suggesting that the reflection component varies less strongly than the
primary power-law over the course of the observation.
To verify this, however, we would need to observe NGC 3516 over a larger 
X-ray flux range than in the current {\it Suzaku} observation to potentially observe any 
significant variability in the PIN band.

Finally, we discuss the origin of the material that gives rise to the observed 
Compton reflection hump. The primary and high-ionization absorbers lack the necessary column 
density, and are excluded. We next consider an origin in the same material that yields 
either the broad or narrow Fe lines.
George, Fabian (1991) calculated that $R$ = 1 corresponds to an observed Fe K$\alpha$ line $EW$ (relative to 
a primary continuum with a photon index of 1.9) of 140 eV for neutral Fe, assuming an inclination
angle of 25$\arcdeg$. However, George, Fabian (1991) used the elemental
abundances of Morrison, McCammon (1983), where the Fe number abundance relative to hydrogen
was $A_{\rm Fe}$ = 3.3 $\times$ 10$^{-5}$. More recent papers have slightly lower values of 
$A_{\rm Fe}$, 2.7--3.0 $\times$ 10$^{-5}$ (Lodders 2003; Wilms et al.\ 2000).
The expected equivalent width corresponding to $R$ = 1 is thus 115--125 eV.
In our baseline model, we found a best-fit value of $R$ = 1.7, which corresponds 
to an expected line $EW$ (relative to the
primary continuum) of 200--215 eV, a value in between the observed $EW$s of the broad line (287 eV 
in the baseline model) and the narrow line (123 eV). 
It is thus not clear from this measurement alone whether the total
Compton reflection continuum is associated with the broad line (disk), narrow line (a distant origin), or both. 
That is, while is it a possibility that at least some portion
of the Compton reflection component is associated with the broad Fe K component,
we cannot exclude the possibility that the narrow line contributes as well
and that there is reflection off cold, distant material.
For example, in $\S$3.3, we demonstrated that a model wherein
there existed both blurred reflection from an ionized disk plus reflection 
from cold, distant material, such as the molecular torus, provided a good fit to the data.
In addition, we demonstrated in this section that
the observed $EW$ of the narrow Fe K line means we cannot rule out a contribution to the narrow Fe line, 
and to the reflection continuum as well, from Compton-thick material out of the line of sight. 



\section{Summary of Main Results}

We have reported on a 150 ksec observation of NGC 3516 obtained with the {\it Suzaku} 
observatory in October 2005. The good exposure times after screening were 135 ks 
for each of the XIS cameras and 106 ks for the HXD-PIN. 

Our best-fit broadband model included a primary power-law with photon index 
$\Gamma$ = 1.904$\pm$0.025 in our baseline model, absorbed by a column of material 
with $N_{\rm H}$ = 5.5$\pm$0.2 $\times$ 10$^{22}$ cm$^{-2}$ and with log($\xi$) = 0.3 
erg cm s$^{-1}$. We modeled the soft band continuum emission using a power-law component 
which could represent
nuclear emission off optically-thin material, unobscured ``leaked'' nuclear 
emission, or a blend of both. The hard X-ray absorber could thus be 
a partial coverer, with a covering fraction $>$96$\%$, or it could obscure the
X-ray continuum source completely. If this absorber is the 
same ``heavy'' absorption component reported by Turner et al.\ (2005) in the 2001 
{\it XMM-Newton} observations,  then between 2001 and 2005 the column density 
of this absorber decreased by a factor of 4.5, while the covering fraction 
increased substantially, leading to an observed 0.5--2.0 keV flux a factor of 
2--3 lower in 2005 than in 2001. Subsequently, by the 2006 October {\it XMM-Newton} observation,
the covering fraction returned to approximately the same
level observed in 2001.
One possibility for the variations in the properties of the obscuring material between the 2001, 2005, and 2006 observations
could be the presence of discrete clouds or filaments within a few light years of the black hole
traversing the line of sight; the equatorial molecular torus invoked in Seyfert unification schemes 
is likely not directly responsible.
We also modeled a highly-ionized absorber with 
a column density $N_{\rm H}$ of 4.0$^{+4.6}_{-3.1}$ $\times$ 10$^{22}$ cm$^{-2}$, 
ionization parameter log($\xi$) = 3.7$^{+0.3}_{-0.7}$ erg cm s$^{-1}$, assumed 
to have a covering fraction of unity.

Our baseline model also included a dozen narrow K-shell emission lines originating in 
He-like N, O, Ne and Mg, H-like C, N, O, Ne and Mg and three RRC features, 
consistent with an origin in photo-ionized material.
However, we cannot exclude a contribution from collisionally-ionized material, as 
suggested by the presence of an Fe L-shell {\sc XVII} line near 0.83 keV.


The broad Fe K$\alpha$ line has been robustly detected:  
we can distinguish between the
curvature in the observed continuum due to a partial coverer and that due to 
a broad diskline; we conclude that for this observation of NGC 3516, a diskline component
is required and that neither a cold nor ionized partial coverer can mimic the continuum curvature associated with the diskline
component. The broad and narrow lines are decoupled (detected independently) at high significance,
thanks to the narrow response of the XIS and the subsequent high signal/noise ratio in the narrow line.
In our best-fit model, we find the Compton reflection strength to be
the value of $R$ = 1.7$^{+0.4}_{-0.5}$.
The narrow Fe K$\alpha$ line, meanwhile, has a FWHM velocity width of $<$4900 km s$^{-1}$ (99$\%$ confidence
level for two interesting parameters), consistent with an 
origin in material with the same velocity as NGC 3516's BLR, though a contribution from
material with lower velocity widths cannot be excluded. 
It is possible that the primary hard X-ray absorber may also be responsible for
emitting the narrow Fe K$\alpha$ line, though there may be a  
contribution from material lying out of the line of sight, such the putative molecular torus.

\vspace{+0.5cm}
The authors gratefully acknowledge the dedication and hard 
work of the {\it Suzaku} hardware teams and operations staff for making this 
observation possible and for assistance with data calibration and analysis. 
This research has made use of HEASARC online services, supported by NASA/GSFC. 
This research has also made use of the NASA/IPAC Extragalactic Database,
operated by JPL/California Institute of Technology, under contract with NASA.

 \clearpage


\begin{longtable}{lcccc}
  \caption{Soft X-ray Emission Lines}\label{tab1}
  \hline              
Line           &  Line          & Intensity                        &  $EW$  &                 \\ 
Energy (keV)   & Identification & (10$^{-6}$ ph cm$^{-2}$ s${-1}$) &  (eV)  & --$\Delta\chi^2$ \\ 
\endhead
\multicolumn{5}{l}{Results are for our best-fit baseline model.} \\
\multicolumn{5}{l}{$^{\dagger}$ denotes a fixed parameter.}\\
\endfoot
  \hline
0.366$^{\dagger}$         & C {\sc VI}               & 88$\pm26$              & 32$\pm$9              &  22.71 \\ 
0.427$^{\dagger}$         & N {\sc VI}               & 42$\pm$11              & 20$\pm$5              &  25.00 \\ 
0.495$\pm$0.005  & N {\sc VII}/C {\sc VI} RRC blend  & 34$\pm$7               & 22$\pm$5              &  47.34 \\ 
0.563$\pm$0.004           & O {\sc VII}              & 77$\pm$10              & 63$\pm$8              & 154.92 \\ 
0.665$^{+0.012}_{-0.008}$ & O {\sc VIII}             & 15$\pm$4               & 16$\pm$4              &  30.04 \\ 
0.739$^{\dagger}$         & O {\sc VII} RRC          &  6.3$\pm$2.9           & 8.6$\pm$4.0           &  12.62 \\ 
0.830$\pm$0.017           & Fe L {\sc XVII}          &  6.2$^{+7.8}_{-2.7}$   & 8.4$^{+10.6}_{-3.7}$  &  14.36 \\ 
0.871$^{\dagger}$         & O {\sc VIII} RRC         & 13$^{+5}_{-10}$        & 20$^{+8}_{-15}$       &   6.94 \\ 
0.917$\pm$0.007           & Ne {\sc IX}              & 16$^{+4}_{-3}$         & 26$^{+7}_{-5}$        & 141.41 \\ 
1.020$^{\dagger}$         & Ne {\sc X}               &  5.6$\pm$1.5           & 14$\pm$4              &  37.45 \\ 
1.196$^{\dagger}$         & Ne {\sc IX} RRC          &  5.2$\pm$1.2           & 18$\pm$4              &  48.22 \\ 
1.351$\pm$0.006          & Mg {\sc XI}              &  6.9$^{+0.9}_{-1.3}$   & 30$^{+4}_{-6}$         &  103.62 \\ 
1.471$^{\dagger}$         & Mg {\sc XII}             &  2.8$\pm$1.0           & 14$\pm$5              &  20.25 \\ 
\hline 
\end{longtable}


\begin{longtable}{llcc}
  \caption{Best-fit parameters for the 1-absorber model and the baseline (2-absorber) model\label{tab2}}
  \hline 
    &  & 1-absorber Model & Baseline (2-absorber) Model  \\ 
\endhead
\multicolumn{4}{l}{$^a$Units of power-law normalization are ph cm$^{-2}$ s$^{-1}$ keV$^{-1}$ at 1 keV.} \\ 
\multicolumn{4}{l}{$^b$ The inclination angle of the Compton reflection component (modeled with {\sc pexrav}) was set at 20$\arcdeg$.} \\ 
\endfoot
  \hline
\multicolumn{2}{l}{$\chi^2$/$dof$}               & 1521.46/1419                       & 1485.04/1417         \\
\multicolumn{2}{l}{$\Gamma$}                     & 1.867$^{+0.033}_{-0.042}$          & 1.904$\pm$0.025        \\
\multicolumn{2}{l}{Primary power-law norm.$^a$}  & 8.8$^{+0.4}_{-0.2} \times 10^{-3}$ & 9.5$\pm$0.5 $\times 10^{-3}$        \\
Primary absorber            & $N_{\rm H}$ (10$^{22}$ cm$^{-2}$)  & 5.5$\pm$0.2                        & 5.5$\pm$0.2          \\       
                            & log($\xi$) (erg cm s$^{-1}$)       & 0.3$\pm$0.1                        & 0.3$\pm$0.1         \\
\multicolumn{2}{l}{Soft power-law norm.$^a$   } & 3.9$\pm$0.2 $\times 10^{-4}$       & 4.0$\pm$0.2 $\times 10^{-4}$        \\
High-ionization absorber   & $N_{\rm H}$ (10$^{22}$ cm$^{-2}$)  &     --                             & 4.0$^{+4.6}_{-3.1}$        \\
                            & log($\xi$) (erg cm s$^{-1}$)       &     --                             & 3.7$^{+0.3}_{-0.7}$        \\
Fe K$\alpha$ diskline       & Energy (keV)                       &  6.40$^{+0.18}_{-0.00}$            & 6.48$^{+0.13}_{-0.08}$         \\
                            & $R_{\rm in}$ ($R_{\rm g}$)         &  $<$3.5                            & $<$3.5         \\
                            & Inclination $i$ ($\arcdeg$)        &  $<$23                               & 25$^{+7}_{-8}$                 \\
                            & Intensity (ph cm$^{-2}$ s$^{-1}$)  &  1.33$^{+0.21}_{-0.16}$ $\times 10^{-4}$   & 1.22$^{+0.21}_{-0.10}$ $\times 10^{-4}$        \\
                            &  $EW$ (eV)                         &  296$^{+47}_{-36}$                 & 287$^{+49}_{-24}$        \\
Narrow Fe K$\alpha$ line    & Intensity (ph cm$^{-2}$ s$^{-1}$)  &  5.3$\pm$0.4 $\times 10^{-5}$      &   5.2$\pm$0.3 $\times 10^{-5}$  \\
                            & $EW$ (eV)                          &  132$\pm$10                       & 123$\pm$7 \\
\multicolumn{2}{l}{Compton reflection strength $R$}              &  1.4$^{+0.2}_{-0.3}$$^b$               & 1.7$^{+0.4}_{-0.5}$         \\
\hline
\end{longtable}


\begin{longtable}{llccc}
  \caption{Best-fit parameters for models with partial covering components\label{tab3}}
  \hline 
&  & Model PC1  & Model PC2 & Model PC3 \\
\endhead
\multicolumn{5}{l}{An asterisk (*) denotes that the parameter uncertainty pegged at the hard limit.} \\
\multicolumn{5}{l}{$^a$ Units of power-law normalization are ph cm$^{-2}$ s$^{-1}$ keV$^{-1}$ at 1 keV.} \\
\multicolumn{5}{l}{$^b$ parameters tied to those of the primary absorber.}  \\ 
\multicolumn{5}{l}{$^c$ The inclination angle of the Compton reflection component (modeled with {\sc pexrav}) was set at 25$\arcdeg$.}\\
\endfoot
  \hline
\multicolumn{2}{l}{$\chi^2$/$dof$}      &  1485.69/1416                
                                        &  1480.48/1414      
                                        &  1676.57/1418      \\
\multicolumn{2}{l}{$\Gamma$}                              & 1.931$^{+0.027}_{-0.057}$    
                                                          & 1.972$^{+0.011}_{-0.036}$  
                                                          & 2.051$^{+0.046}_{-0.048}$       \\
\multicolumn{2}{l}{Primary power-law norm.$^a$ }  & 7.5$\pm$0.2 $\times 10^{-3}$    
                                                  & 9.8$^{+0.3}_{-0.7}$ $\times 10^{-3}$   
                                                  & 1.3$^{+0.1}_{-0.2}$ $\times 10^{-2}$ \\
Primary absorber            & $N_{\rm H}$ (10$^{22}$ cm$^{-2}$)  & 5.5$\pm$0.1   & 6.1$^{+0.2}_{-0.1}$  & 5.3$\pm$0.2      \\
                            & log($\xi$) (erg cm s$^{-1}$)       & 0.3$\pm$0.1  &  0.3$\pm$0.1    & 0.1$^{+0.2}_{-0.1}$      \\
Partial covering component  & $N_{\rm H,PC}$ (10$^{22}$ cm$^{-2}$)  & 5.5$^b$                      & 1.0$^{+1.3}_{-0.2}$   & 7.8$^{+68}_{-3.8*}$      \\
                            & log($\xi_{\rm PC}$) (erg cm s$^{-1}$)       & 0.3$^b$                      &  --1.3$^{+0.5}_{-1.7*}$  & --0.4$^{+1.2}_{-2.6*}$      \\
                            & Power-law Norm.$^a$                & 2.4$\pm$0.2 $\times 10^{-3}$ &  8.8$^{+3.9}_{-2.3}$ $\times 10^{-4}$  & $<$2.4 $\times$ 10$^{-3}$      \\
\multicolumn{2}{l}{Soft power-law norm.$^a$   } & 4.0$\pm$0.1 $\times 10^{-4}$ &  3.7$^{+0.2}_{-0.3}$ $\times 10^{-4}$ & 4.1$^{+0.1}_{-0.3}$ $\times 10^{-4}$      \\
High-ionization absorber    & $N_{\rm H}$ (10$^{22}$ cm$^{-2}$)  & 4.1$\pm$2.5                    & 3.3$^{+1.8}_{-1.3}$        & 11.7$^{+2.8}_{-0.4}$     \\
                            & log($\xi$) (erg cm s$^{-1}$)       &    3.3$^{+0.3}_{-0.1}$         & 3.5$^{+0.8}_{-0.2}$       & 3.2$\pm$0.1      \\
Fe K$\alpha$ diskline       & Energy (keV)                       & 6.45$^{+0.15}_{-0.05*}$         & 6.45$^{+0.10}_{-0.05}$    &  --      \\
                            & $R_{\rm in}$ ($R_{\rm g}$)         & $<$3.2                          & $<$4.3         &  --    \\
                            & Inclination $i$ ($\arcdeg$)        &  26$^{+4}_{-6}$                  & 26$^{+4}_{-8}$             &  --      \\
                            & Intensity (ph cm$^{-2}$ s$^{-1}$)  &  1.18$^{+0.25}_{-0.17}$ $\times 10^{-4}$ &  1.13$^{+0.25}_{-0.15}$ $\times$ 10$^{-4}$ & --      \\
                            & $EW$  (eV)                         &  268$^{+57}_{-39}$               & 266$^{+56}_{-34}$  &     --        \\
Narrow Fe K$\alpha$ line    & Intensity (ph cm$^{-2}$ s$^{-1}$)  &   5.4$^{+0.5}_{-0.4}$ $\times 10^{-5}$  &  5.4$^{+0.5}_{-0.4}$ $\times 10^{-5}$  & 8.8$\pm$0.7 $\times 10^{-5}$  \\
                            & $EW$ (eV)                          &  153$^{+14}_{-11}$    &   134$^{+12}_{-10}$  &  226$\pm$18  \\
\multicolumn{2}{l}{Compton reflection strength $R$}              &  1.7$^{+0.2}_{-0.4}$            & 1.8$^{+0.4}_{-0.5}$    & 2.4$^{+0.9}_{-0.3}$$^c$       \\    
\hline
\end{longtable}


\begin{longtable}{lccc}
\caption{Comparison to the 2001 {\it XMM-Newton} observations\label{tab4}}
  \hline 
  &  {\it Suzaku} & {\it XMM-Newton}   & {\it XMM-Newton} \\
  &  Oct.\ 2005    & Apr.\ 2001         & Nov.\ 2001    \\
\endhead
\multicolumn{4}{l}{Results for the {\it XMM-Newton} spectra were taken from Turner et al.\ (2005).}\\
\multicolumn{4}{l}{Results for {\it Suzaku} are from the best-fit baseline model. }\\
\multicolumn{4}{l}{$^{\dagger}$ denotes a fixed parameter.} \\ 
\endfoot
\hline
Absorption-corrected 2--10 keV flux  (erg cm$^{-2}$ s$^{-1}$)   &  3.4 $\times$ 10$^{-11}$   & 2.7 $\times$ 10$^{-11}$   & 1.9 $\times$ 10$^{-11}$ \\
Observed 0.5--2.0 keV flux           (erg cm$^{-2}$ s$^{-1}$)   &  1.3 $\times$ 10$^{-12}$   & 4.3 $\times$ 10$^{-12}$   & 2.9 $\times$ 10$^{-12}$ \\
Primary absorber $N_{\rm H}$  (10$^{22}$ cm$^{-2}$)             &  5.5$\pm$0.2             & 25$\pm$1                &  25$\pm$1   \\
Primary absorber covering fraction                              & 96--100$\%$              &   44$\pm$6$\%$          &  58$\pm$5$\%$    \\
High-ionization $N_{\rm H}$   (10$^{22}$ cm$^{-2}$)             & 4.0$^{+4.6}_{-3.1}$      &  1.6$^{\dagger}$        &  1.6$^{\dagger}$ \\
\hline
\end{longtable}


\clearpage

\begin{figure}
\begin{center}
\FigureFile(80mm,120mm){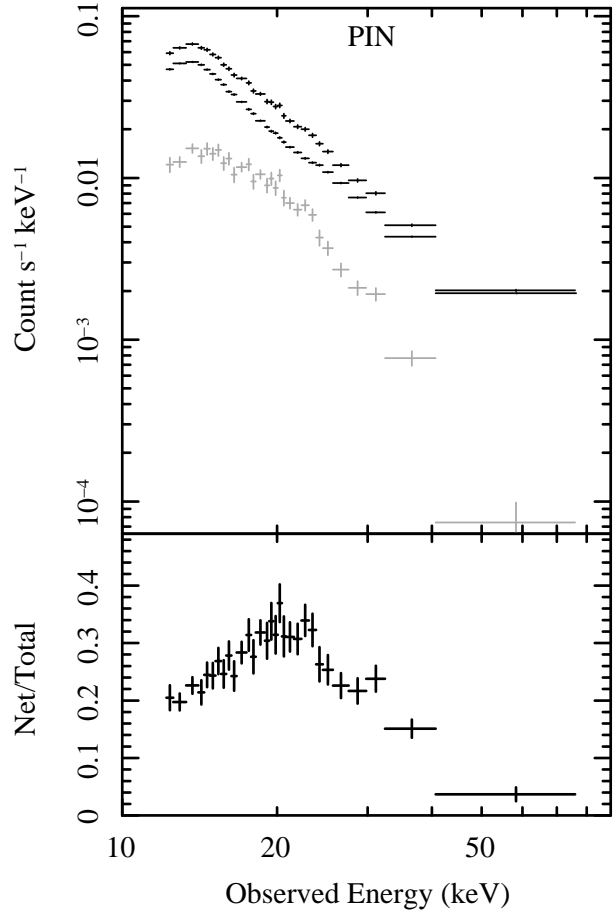}
\end{center}
\caption{HXD-PIN spectrum. The upper panel shows
the net source spectrum ({\it gray points}), the background 
({\it lower black points}),
and the total (source + background) spectrum ({\it upper black points}).
The PIN spectra have been binned such that the net spectrum has a minimum
signal-to-noise ratio of 8$\sigma$ per bin.
The lower panel shows the ratio of the net source spectrum to
the total spectrum.}\label{fig1}
\end{figure}

\begin{figure}
\begin{center}
\FigureFile(90mm,155mm){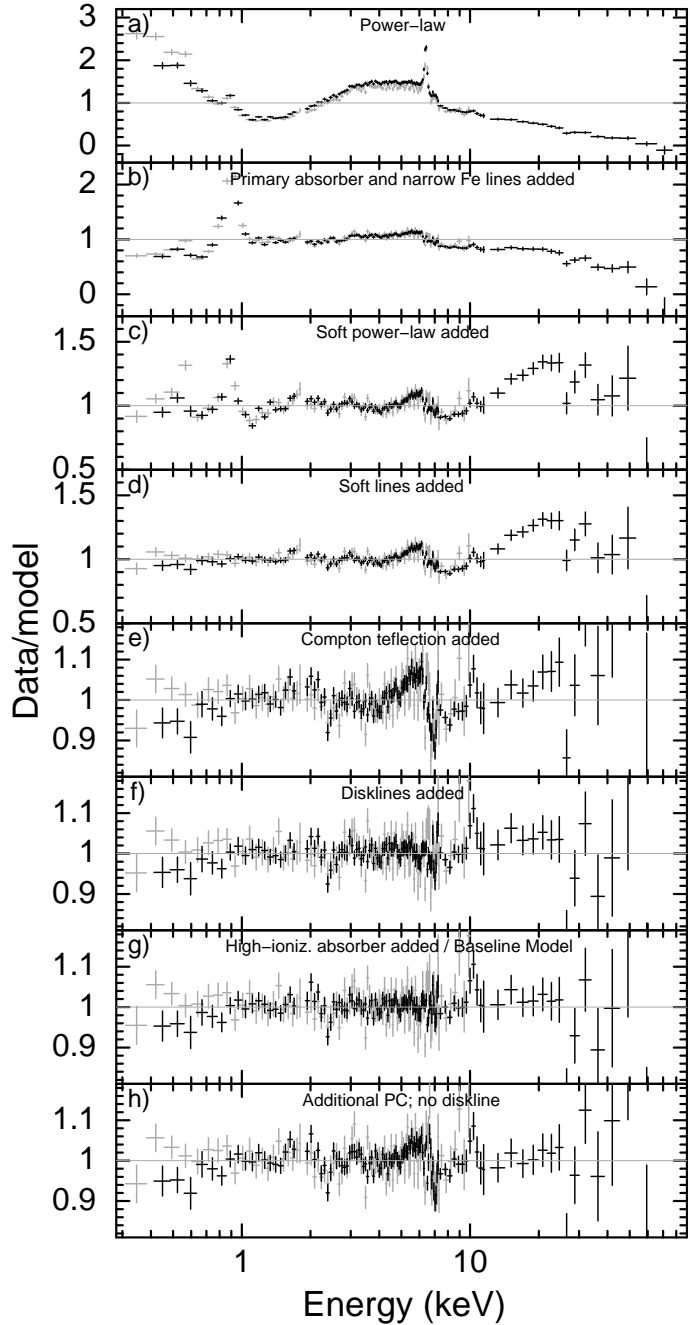}
\end{center}
\caption{Data/model residuals for selected broadband models.
Panel a) shows residuals to a simple power-law.
Panel b) shows the residuals after the primary absorber
and narrow Fe lines are included in the model.
Panel c) shows residuals after the soft power-law is included.
In panel d), the soft emission lines have been modeled.
In panel e), a Compton reflection component has been
added (using {\sc pexrav} in {\sc xspec}). In panel f), diskline components have been added.
In panel g), the high-ionization absorber has been added to
yield our baseline model.
Panel h) shows the results when the broad Fe lines are removed
and substituted with an additional partial covering component.
Black points $<$12 keV denote the (co-added) XIS-FI
spectrum. Gray points denote the XIS-BI spectrum.
Black points $>$12 keV denote the PIN data.
Rest-frame energies are shown. All data have been rebinned with
a binning factor of 5, though XIS data $>$7.2 keV have been rebinned
by a factor of 25 for clarity.}\label{fig2}
\end{figure}

\begin{figure}
\begin{center}
\FigureFile(136mm,104mm){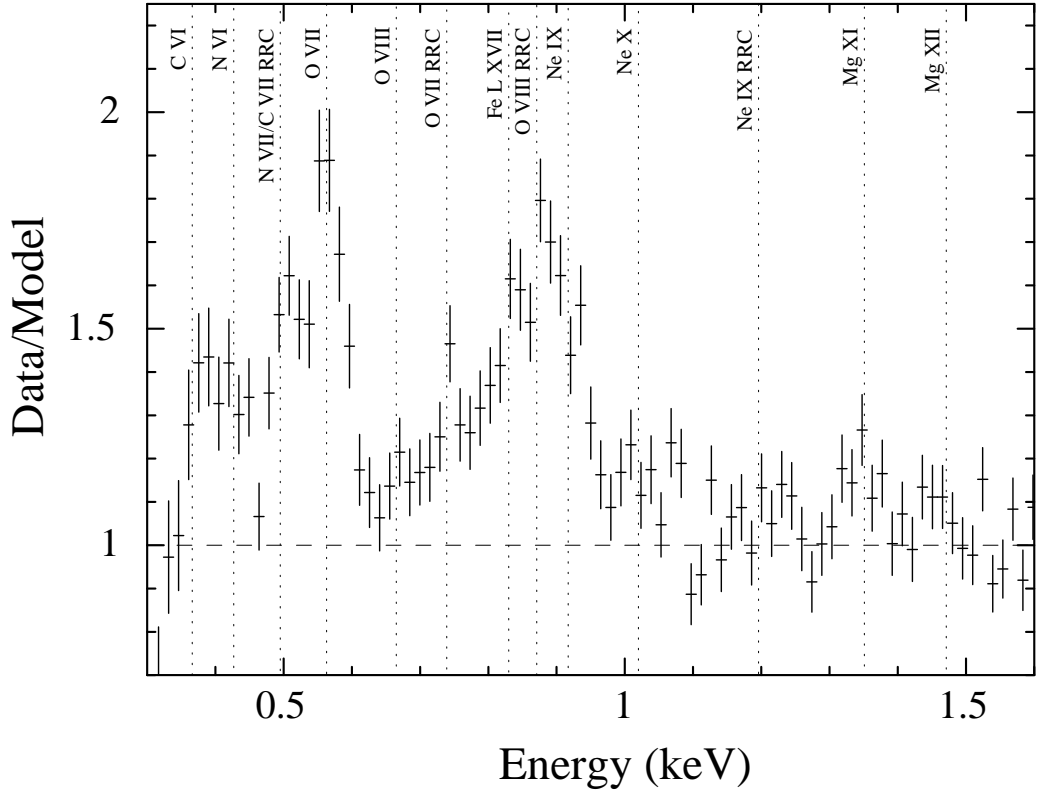}
\end{center}
\caption{Ratio of the soft band data to a simple power-law, 
showing the prominent emission lines.
For clarity, only the XIS BI spectrum is shown.
Rest-frame energies are shown.}\label{fig3}
\end{figure}

\clearpage

\begin{figure}
\begin{center}
\FigureFile(108mm,116mm){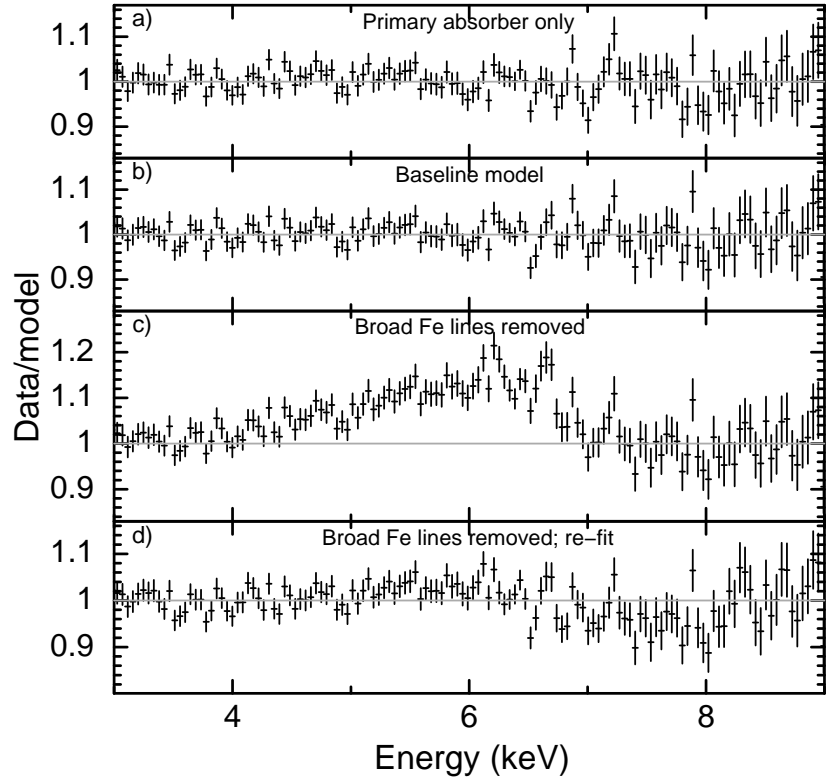}
\end{center}
\caption{Data/model residuals to various models are shown, focusing
on the Fe K bandpass. Data have been rebinned by a factor of 3.
Panel a) shows the residuals when the primary absorber
is the only zone of absorption modeled.
Panel b) shows the residuals to the baseline model, with two
zones of absorption; note that residuals near 6.9 keV are
now improved.
In panel c), the broad Fe lines have been removed from
the baseline model. When the model is then refit,
correlated data/model residuals appear in the Fe K bandpass, as illustrated in
panel d).}\label{fig4}
\end{figure}

\clearpage

\begin{figure}
\begin{center}
\FigureFile(133mm,85mm){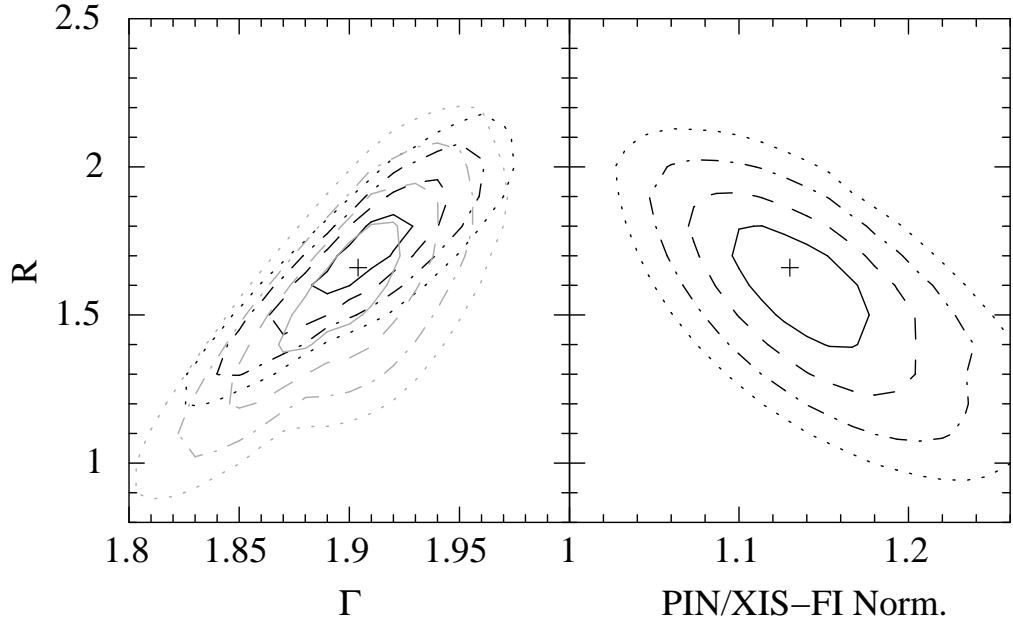}
\end{center}
\caption{{\it Left:} Contour plot of the Compton reflection fraction $R$
versus the photon index of the primary power-law.
The black contours show results when 
the PIN/XIS-FI normalization was kept fixed at 1.13.
The gray contours show results when the PIN/XIS-FI normalization was 
left free.
{\it Right:} Contour plot of $R$ versus the PIN/XIS-FI normalization;
$\Gamma$ was a free parameter. In both plots, solid, 
dashed, dot-dashed, and dotted lines denote 68$\%$, 95.4$\%$, 99.73$\%$, and 99.99$\%$
confidence levels, respectively.}\label{fig5}
\end{figure}

\clearpage

\begin{figure}
\begin{center}
\FigureFile(80mm,50mm){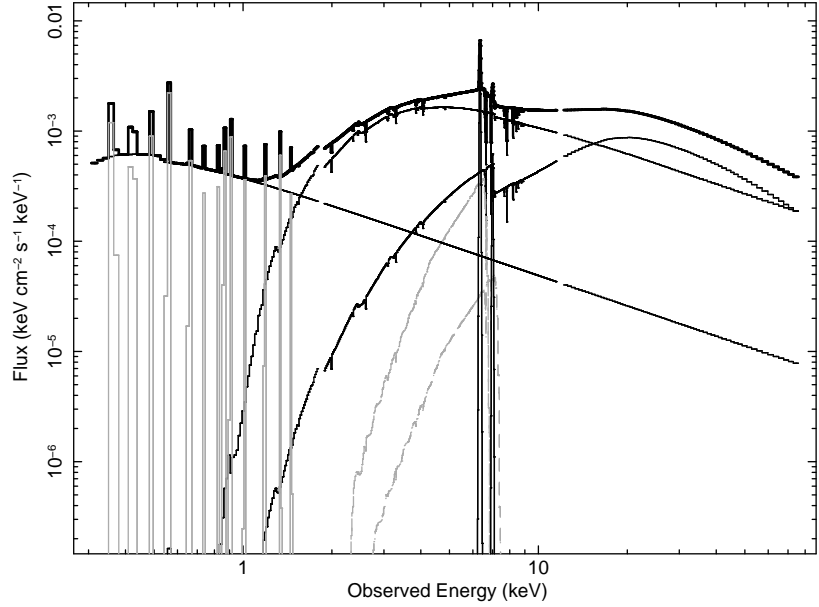}
\end{center}
\caption{Unfolded broadband spectrum for the baseline model.
The thick black line is the total spectrum.
Thin black lines denote the primary, absorbed power-law,
the soft power-law, and the reflection component peaking
at 20 keV. Gray dashed lines denote the narrow Fe lines.
Gray solid lines in the Fe K bandpass are the diskline components.
Gray solid lines in the soft band denote the
soft emission lines.}\label{fig6}
\end{figure}


\begin{figure}
\begin{center}
\FigureFile(107mm,108mm){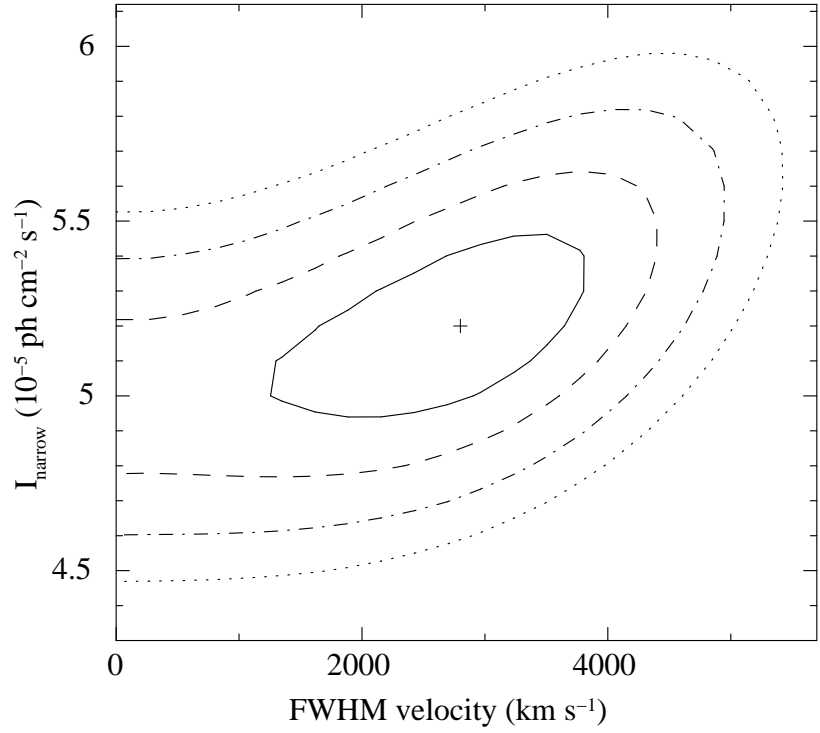}
\end{center}
\caption{Contour plot showing the
intensity of the narrow K$\alpha$ line versus the FWHM velocity width.
68$\%$, 95.4$\%$, 99.73$\%$ and 99.99$\%$ confidence levels for two interesting parameters are plotted
(solid, dashed, dot-dashed, and dotted lines, respectively).}\label{fig7}
\end{figure}


\begin{figure}
\begin{center}
\FigureFile(107mm,18mm){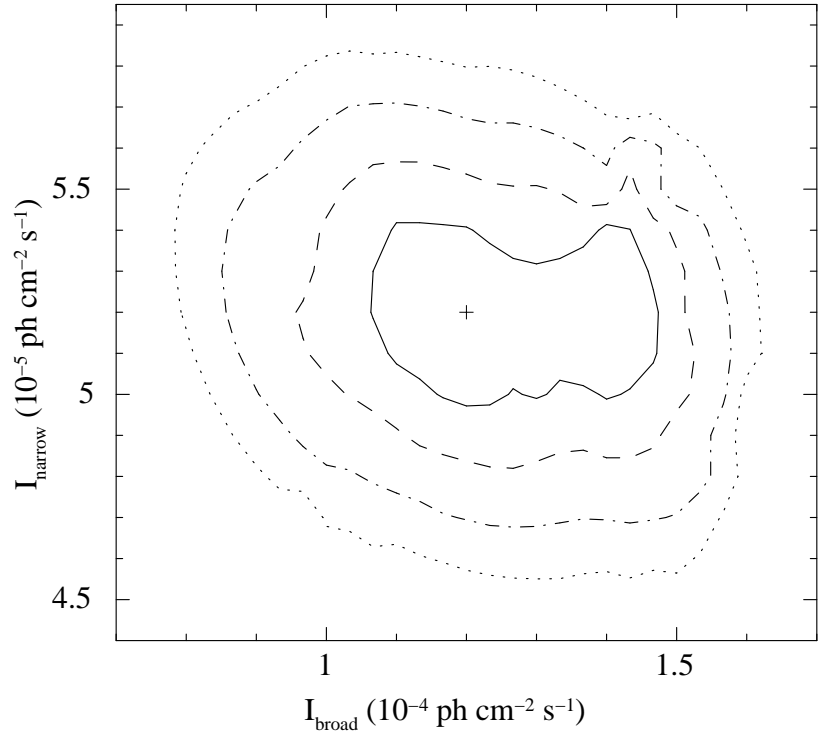}
\end{center}
\caption{Contour plot showing the
intensity of the narrow K$\alpha$ line versus that of the
broad K$\alpha$ line.
68$\%$, 95.4$\%$, 99.73$\%$ and 99.99$\%$ confidence levels for two interesting parameters are plotted
(solid, dashed, dot-dashed, and dotted lines, respectively).
Note that the 99.99$\%$ confidence level contours do not hit 0, 
demonstrating that the lines are decoupled
at the 4$\sigma$ level.}\label{fig8}
\end{figure}


\begin{figure}
\begin{center}
\FigureFile(115mm,116mm){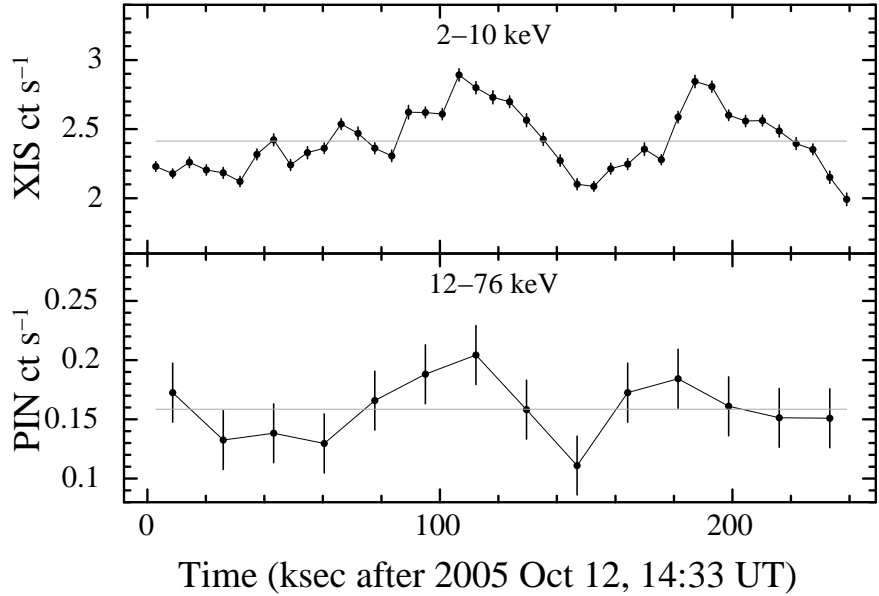}
\end{center}
\caption{Orbitally binned, background-subtracted 
2--10 keV XIS light curve, summed over all four XIS cameras
and background subtracted {\it (top)},
and background-subtracted
12--76 keV PIN light curve, binned to every three orbits
{\it (bottom)}.
The error bars show only the statistical error of the count rates;
systematic uncertainty associated with subtraction of the
non-X-ray background in the PIN is roughly 0.01 ct s$^{-1}$.
The mean count rates are 2.41 ct s$^{-1}$ for 2--10 keV
and 0.16  ct s$^{-1}$ for 12--76 keV.}\label{fig9}
\end{figure}

\begin{figure}
\begin{center}
\FigureFile(125mm,75mm){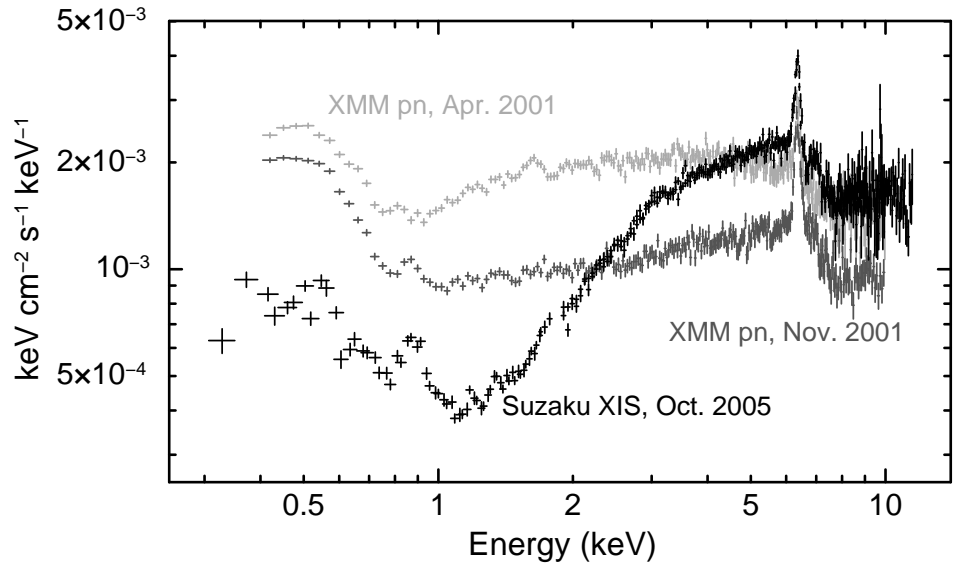}
\end{center}
\caption{Unfolded spectra for the {\it Suzaku} and 2001 {\it XMM-Newton} observations.
The black points denote the {\it Suzaku} XIS-FI and BI spectra, rebinned by a factor of 3.
The light and dark gray points denote the
{\it XMM-Newton} EPIC-pn spectra for April 2001 and November 2001, respectively;
data are rebinned by a factor of 6.}\label{fig10}
\end{figure}

\end{document}